%% file: author/0_main.tex
\documentclass[graybox]{svmult}

\usepackage{physics}
\usepackage{type1cm}        
\usepackage{makeidx}         
\usepackage{graphicx}        
\usepackage{multicol}        
\usepackage[bottom]{footmisc}

\usepackage[utf8]{inputenc}
\usepackage{newtxtext}       %
\usepackage[varvw]{newtxmath}       
\usepackage{soul}
\usepackage{tabularx}
\usepackage{array}
\usepackage{enumitem}

\makeindex             

\begin{document}

\title*{Design Automation in Quantum Error Correction}
\author{Archisman Ghosh\orcidID{0000-0002-0264-6687}, Avimita Chatterjee\orcidID{0009-0001-7421-9334} and Swaroop Ghosh\orcidID{0000-0001-8753-490X}}
\institute{Archisman Ghosh \at Pennsylvania State University, University Park, PA, USA \email{apg6127@psu.edu}
\and Avimita Chatterjee \at Pennsylvania State University, University Park, PA, USA \email{amc8313@psu.edu}
\and Swaroop Ghosh \at Pennsylvania State University, University Park, PA, USA \email{szg212@psu.edu}}
%
%
\maketitle


\abstract{Quantum error correction (QEC) underpins practical fault-tolerant quantum computing (FTQC) by addressing the fragility of quantum states and mitigating decoherence-induced errors. As quantum devices scale, integrating robust QEC protocols is imperative to suppress logical error rates below threshold and ensure reliable operation, though current frameworks suffer from substantial qubit overheads and hardware inefficiencies. Design automation in the QEC flow is thus critical, enabling automated synthesis, transpilation, layout, and verification of error-corrected circuits to reduce qubit footprints and push fault-tolerance margins.
This chapter presents a comprehensive treatment of design automation in QEC, structured into four main sections. The first section delves into the theoretical aspects of QEC, covering logical versus physical qubit representations, stabilizer code construction, and error syndrome extraction mechanisms. In the second section, we outline the QEC design flow, detailing the areas highlighting the need for design automation. The third section surveys recent advancements in design automation techniques, including algorithmic $T$-gate optimization, modified surface code architecture to incorporate lesser qubit overhead, and machine-learning-based decoder automation. The final section examines near-term FTQC architectures, integrating automated QEC pipelines into scalable hardware platforms and discussing end-to-end verification methodologies.
Each section is complemented by case studies of recent research works, illustrating practical implementations and performance trade-offs. Collectively, this chapter aims to equip readers with a holistic understanding of design automation in QEC system design in the fault-tolerant landscape of quantum computing.
}

\input{author/section1}

\input{author/section2}
\input{author/section3}

\input{author/section4}

\input{author/section5}

\begin{acknowledgement}
The work is supported in parts by NSF (CNS-2129675, CCF-2210963), gifts from Intel and IBM Quantum Credits.
\end{acknowledgement}
\ethics{Competing Interests}{
The authors have no conflicts of interest to declare that are relevant to the content of this chapter.}


\bibliographystyle{IEEEtran} 
\bibliography{author/refs}
\end{document}

%% file: author/section1.tex
\section{Introduction}
\label{sec:1}

Quantum error correction (QEC) is a foundational component in the realization of fault-tolerant quantum computing (FTQC) \cite{dummies}. Unlike classical bits, quantum bits (qubits) are inherently fragile and susceptible to a wide range of noise processes, including bit-flip, phase-flip, decoherence, and crosstalk errors \cite{noise}. These challenges are exacerbated in the Noisy Intermediate-Scale Quantum (NISQ) era \cite{Preskill_2018}, where quantum hardware is still in its nascent stages of scalability and stability. QEC techniques aim to protect logical quantum information by encoding it into entangled states of multiple physical qubits, thereby enabling detection and correction of errors without direct measurement of the quantum state.
While the theoretical foundations of QEC, such as stabilizer codes \cite{gottesman1997stabilizercodesquantumerror} and surface codes \cite{Fowler_2012}, are well developed, practical implementation remains resource-intensive and operationally complex. This has motivated a growing interest in automating QEC workflows, from circuit-level optimization and code selection to decoder design and syndrome-based correction. As quantum systems scale, automation becomes not merely beneficial but essential to ensure the efficiency, scalability, and adaptability of QEC strategies in modern quantum architectures. 
\subsection{What is Quantum Error Correction?}
QEC essentially protects quantum information against decoherence, gate infidelities, and measurement errors by encoding logical qubits into entangled states of multiple physical qubits \cite{dummies}. This is essential because quantum states are inherently fragile and susceptible to both bit-flip ($X$) and phase-flip ($Z$) errors, which cannot be mitigated by classical redundancy due to the no-cloning theorem \cite{Wootters1982}. QEC employs syndrome measurement and conditional correction without disturbing the encoded state, enabling reliable quantum operations \cite{dummies}. As we scale beyond the NISQ era, QEC becomes indispensable for achieving large-scale FTQC.

\subsubsection*{Why does classical error correction not work?}
Classical error correction codes (ECC), such as Hamming \cite{Hamming1950}, Bose–Chaudhuri–Hoc\-quenghem (BCH) \cite{bch}, and Reed-Solomon codes \cite{rscode}, function by introducing redundancy through the \emph{duplication} of classical bits and performing majority voting or algebraic checks to detect and correct transmission errors. These codes rely on two critical assumptions: (1) data can be freely copied, and (2) data can be measured repeatedly without altering the underlying state. Both assumptions fail in the quantum domain due to fundamental principles of quantum mechanics.
\begin{itemize}[label={--}]
    \item \textbf{No-Cloning Theorem: }A central principle of quantum theory is the no-cloning theorem, which states that it is impossible to create an independent and identical copy of an arbitrary unknown quantum state. Mathematically, there exists no unitary operation $U$ such that $U(\ket{\psi}\otimes\ket{0})=\ket{\psi}\otimes\ket{\psi}, \forall \ket{\psi}\in \mathcal{H}$; $\mathcal{H}$ being the the Hilbert space of qubit states \cite{Wootters1982}\cite{jozsa2002strongernocloningtheorem}. This precludes the use of direct repetition codes (e.g., $0\rightarrow 000$) in the quantum setting, thereby invalidating the core mechanism of classical ECC.
    
    \item \textbf{Measurement-induced collapse: }Classical ECCs perform detection and correction by directly reading bit values at the receiver and applying logical inference. However, in quantum mechanics, measurement is a non-unitary, non-reversible operation that collapses the superposition of a qubit state into one of its computational basis states \cite{vonNeumann2018}. For example, if a qubit is in a state $\ket{\psi} = \alpha\ket{0} + \beta\ket{1}$, any standard measurement in the computational basis results in the state irreversibly collapsing to either $\ket{0}$ or $\ket{1}$, with respective probabilities $|\alpha|^2$ and $|\beta|^2$. This destructive property of measurement prevents direct syndrome extraction, as it would annihilate the very quantum information we intend to protect.

    \item \textbf{Complexity of the Quantum Errors: }In classical systems, the error model is typically limited to bit-flip errors (0 $\leftrightarrow$ 1), which can be handled by checking parity or Hamming distance \cite{dummies}. In contrast, quantum systems are subject to a richer error space spanned by the Pauli group $\mathcal{P} = \{I, X, Y, Z\}$. A general error on a qubit can be expressed as a linear combination of Pauli operators: $E = \alpha I+\beta X+\gamma Y+\delta Z$, where $X$(bit-flip), $Z$(phase-flip), and $Y=iXZ$(combined) all act non-trivially on the state. Therefore, a quantum error-correcting code (QECC) must be able to detect and correct not only $X$ and $Z$-type errors but also their simultaneous occurrence, even when occurring coherently.
    Furthermore, errors in quantum systems are not necessarily discrete events but may be continuous or coherent due to interactions with the environment (decoherence), imprecise gate implementations (gate errors), or imperfect readouts (measurement errors). Classical ECC assumes a discrete and probabilistic error model, which is not sufficient to model quantum noise accurately.
\end{itemize}

\subsection{Stabilizer Formalism}

A stabilizer \cite{gottesman1997stabilizercodesquantumerror}\cite{Dauphinais_2024}\cite{dummies} is an operator that preserves a quantum state under its action. Formally, for a quantum state \( |\psi\rangle \), a stabilizer \( g \) satisfies \( g|\psi\rangle = |\psi\rangle \). Stabilizers are elements of the \emph{Pauli group} \( \mathcal{P}_n \), which is defined as the group of \( n \)-fold tensor products of the single-qubit Pauli matrices \( I, X, Y, Z \), including multiplicative factors \( \{\pm1, \pm i\} \). That is, each element of \( \mathcal{P}_n \) is of the form \( i^\alpha P_1 \otimes P_2 \otimes \cdots \otimes P_n \), where each \( P_j \in \{I, X, Y, Z\} \) and \( \alpha \in \{0,1,2,3\} \). The group has cardinality \( 2 \cdot 4^n \), and is closed under multiplication.
A \emph{stabilizer code} is defined by an abelian subgroup \( \mathcal{S} \subset \mathcal{P}_n \) that does not contain \( -I \), and is generated by a set of \( m = n-k \) independent, commuting operators \( G = \{g_1, g_2, \ldots, g_m\} \). The codespace \( \mathcal{C} \) is the subspace of the Hilbert space \( \mathbb{C}^{2^n} \) that is stabilized by all elements of \( \mathcal{S} \), i.e.,
\[
\mathcal{C} = \left\{ |\psi\rangle \in \mathbb{C}^{2^n} \,\middle|\, g_i |\psi\rangle = |\psi\rangle; \forall g_i \in G \right\}
\]
Because each generator constrains the state to a \( +1 \)-eigenspace, and the generators are independent and commuting, the dimension of \( \mathcal{C} \) is \( 2^k \), corresponding to \( k \) logical qubits encoded in \( n \) physical qubits.
To qualify as valid stabilizer generators, the set \( \{g_1, \ldots, g_m\} \) must satisfy three structural rules:
\begin{enumerate}
    \item \textbf{Each generator is a Pauli string}: That is, each \( g_i \) must be an element of the Pauli group \( \mathcal{P}_n \).
    \item \textbf{Commutativity}: All generators must mutually commute, i.e., \( g_i \otimes g_j = g_j \otimes g_i \) for all \( i, j \).
    \item \textbf{Independence}: No generator may be written as a product of others, i.e., there must be no nontrivial subset \( \{g_{i_1}, \ldots, g_{i_r}\} \) such that their product equals another generator in the set.
\end{enumerate}
Logical operations on encoded qubits are performed using \emph{logical Pauli operators}, denoted \( \bar{X}_j, \bar{Z}_j \), which commute with all elements of \( \mathcal{S} \) but are not themselves in \( \mathcal{S} \). These operators act nontrivially on the codespace and are used to implement quantum gates in a fault-tolerant manner. The overall quantum error correcting code is denoted as \( [[n, k, d]] \), where \( d \) is the minimum weight of any Pauli operator that commutes with all elements of \( \mathcal{S} \) but is not in \( \mathcal{S} \), i.e., the minimum weight of a logical error. The stabilizer formalism thus forms both a mathematical and operational framework for defining quantum error-correcting codes.

\subsection{Constructing a Stabilizer Generator}

The construction of stabilizer generators is guided by the principle that an error must anti-commute with at least one stabilizer to be detected. That is, for a stabilizer operator \( S \) and an error operator \( E \in \mathcal{P}_n \), the stabilizer must satisfy \( [S, E] = SE + ES = 0 \). In such a case, the presence of the error flips the eigenvalue of the stabilizer measurement from \( +1 \) to \( -1 \), thereby providing a detectable syndrome. Consequently, to detect bit-flip (Pauli-\(X\)) errors, the stabilizer must involve Pauli-\(Z\) operators; to detect phase-flip (Pauli-\(Z\)) errors, the stabilizer must involve Pauli-\(X\) operators. This design criterion underpins the structural logic of quantum parity-check circuits.

\begin{figure}
    \centering
    \includegraphics[width=1\linewidth]{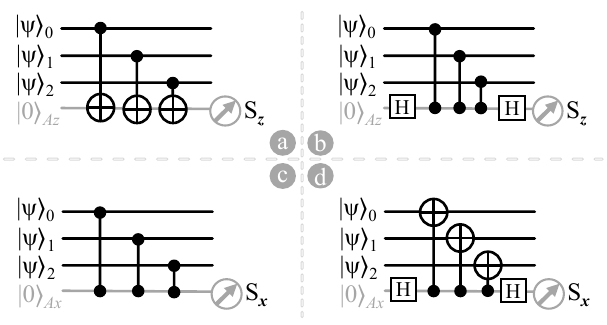}
    \caption{Different approaches to generate a $Z-stabilizer$ and $X-stabilizer$ operating on 3 qubits, namely $\ket{\psi}_0$, $\ket{\psi}_1$ and $\ket{\psi}_2$ \cite{dummies}. The ancilla qubit, on which the stabilizer output is being projected, is represented as $\ket{0}_{Az}$ and $\ket{0}_{Ax}$ for $Z-stabilizers$ and $X-stabilizers$, respectively. The circuits are presented as follows: (a)  This circuit exhibits a $Z-stabilizer$. The ancilla qubit, $\ket{0}_{Az}$ measures $Z\ket{\psi}_0 \otimes Z\ket{\psi}_1 \otimes Z\ket{\psi}_2$, commonly denoted as $Z_0 \otimes Z_1 \otimes Z_2$.
    (b) This circuit demonstrates the same $Z-stabilizer$ as in (a). The stabilizer is regenerated using $H$ gates and $CZ$ gates instead of $CNOT$ gates. The ancilla qubit, $\ket{0}_{Az}$ also measures $Z_0 \otimes Z_1 \otimes Z_2$.
    (c) This circuit represents an $X-stabilizer$. The ancilla qubit, $\ket{0}_{Ax}$ measures $X\ket{\psi}_0 \otimes X\ket{\psi}_1 \otimes X\ket{\psi}_2$, commonly denoted as $X_0 \otimes X_1 \otimes X_2$.
    (d) This circuit portrays the same $X-stabilizer$ as mentioned in (c). The stabilizer is recreated using $H$ gates and $CNOT$ gates instead of $CZ$ gates. The ancilla qubit, $\ket{0}_{Ax}$ also measures $X_0 \otimes X_1 \otimes X_2$.
    }
    \label{fig:all_stab}
\end{figure}

In practical implementations, stabilizers are measured using ancilla-based circuits. Consider a system of three data qubits \( |\psi_0\rangle, |\psi_1\rangle, |\psi_2\rangle \) (Fig. \ref{fig:all_stab}). A typical \(Z\)-type stabilizer generator acts as the parity check \( Z_0 \otimes Z_1 \otimes Z_2 \), which anti-commutes with \(X\)-type errors on any of the three qubits. The syndrome is extracted using an ancilla qubit initialized to \( |0\rangle \), which interacts with each data qubit via a CNOT gate (with data as control, ancilla as target). After the sequence of CNOTs, the ancilla is measured in the computational basis. If no error is present, the measurement yields \(+1\); if a bit-flip error has occurred on one of the data qubits, the ancilla's outcome flips to \(-1\), thus enabling error detection without directly measuring the data qubits.
Detection of \(Z\)-type errors requires an analogous structure using \(X\)-type stabilizers. Since hardware platforms often do not support direct controlled-\(X\) interactions, these stabilizers are constructed by conjugating the circuit with Hadamard gates. Specifically, to measure \( X_0 \otimes X_1 \otimes X_2 \), Hadamard gates are first applied to each data qubit and the ancilla (converting the \(X\)-basis to the \(Z\)-basis), followed by the same CNOT sequence as for a \(Z\)-stabilizer, and finally another layer of Hadamard gates is applied to restore the original basis. This method transforms a parity check in the \(X\)-basis into a standard \(Z\)-basis measurement on the ancilla.

\begin{figure}[t]
    \centering
    \includegraphics[width=1\linewidth]{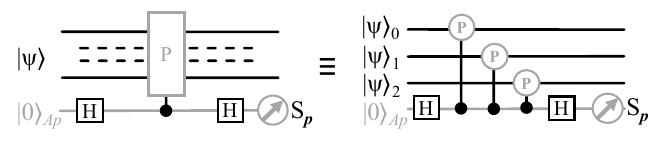}
    \caption{Illustration of a universal circuit of a stabilizer generator, in which $P$ represents a Pauli-gate operation. This circuit measures $P\ket{\psi}_0 \otimes P\ket{\psi}_1 \otimes P\ket{\psi}_2 \equiv P_0 \otimes P_1 \otimes P_2$. Usually for a stabilizer generator, $P \in \{X, Z\}$ \cite{dummies}.}
    \label{fig:stab_gen}
\end{figure}

A general and universal circuit for measuring stabilizer generators of the form \( P_0 \otimes P_1 \otimes \cdots \otimes P_{n-1} \), where \( P \in \{X, Z\} \), uses a sequence of controlled-\(P\) gates mediated by a single ancilla qubit (Fig.\ref{fig:stab_gen}). The circuit begins by preparing the ancilla in the state \( |0\rangle \), followed by a Hadamard gate to place it in the superposition \( (|0\rangle + |1\rangle)/\sqrt{2} \). Then, a sequence of controlled-\(P\) gates is applied such that each control is the ancilla and each target is a data qubit. Finally, a second Hadamard gate is applied to the ancilla before measurement. The resulting quantum state evolves as follows:
\begin{align*}
\text{After first Hadamard:} &\quad \frac{1}{\sqrt{2}} \left( |0\rangle + |1\rangle \right) \otimes |\psi\rangle, \\
\text{After controlled-}P: &\quad \frac{1}{\sqrt{2}} \left( |0\rangle \otimes |\psi\rangle + |1\rangle \otimes P |\psi\rangle \right), \\
\text{After second Hadamard:} &\quad \frac{1}{2} \left( |0\rangle \otimes (I + P)|\psi\rangle + |1\rangle \otimes (I - P)|\psi\rangle \right).
\end{align*}

Measuring the ancilla in the computational basis projects the system onto one of two branches: the outcome \( |0\rangle \) corresponds to the eigenvalue \( +1 \), with probability proportional to \( \| (I + P)|\psi\rangle \|^2 \); the outcome \( |1\rangle \) corresponds to the eigenvalue \( -1 \), indicating that an error has occurred. Notably, if an error commutes with the stabilizer (i.e., \( [P, E] = 0 \)), the measurement outcome remains unaffected; if it anti-commutes, the measurement outcome flips, yielding an eigenvalue of \( -1 \).
This technique ensures that even if the physical error occurs at an arbitrary angle (i.e., not precisely a Pauli operator), the stabilizer circuit projects the state onto a well-defined eigenstate of the stabilizer operator, suppressing coherence across the error subspace, rendering the error syndrome detectable and classically interpretable.

\subsection{Circuit Architecture of a QECC}
Quantum error correction codes (QECCs) are not merely algebraic constructs—they manifest physically as quantum circuits comprising well-defined components that collectively facilitate the encoding, monitoring, and preservation of logical quantum information \cite{Reed_2012}. In this section, we describe the circuit-level structure of a QECC, focusing on the practical instantiation of the encoding and stabilizer modules, while deferring the decoding and correction logic to subsequent sections. A typical quantum error correction circuit can be decomposed into three interdependent stages:
\begin{itemize}[label={--}]
    \item \textbf{Encoding:} The initial logical state, often denoted $|\psi\rangle_k \in \mathcal{H}_{2^k}$, is mapped to a higher-dimensional subspace $|\psi\rangle_L \in \mathcal{H}_{2^n}$, known as the codespace, via a unitary encoding circuit. This stage establishes logical redundancy by entangling the logical qubits with additional physical qubits.

    \item \textbf{Stabilizer Interaction Layer:} A set of ancilla-assisted subcircuits are employed to perform parity checks---each associated with a stabilizer generator. These checks involve entangling gates (typically CNOTs or controlled-Zs) between the data qubits and auxiliary ancilla qubits.

    \item \textbf{Syndrome Projection:} Each ancilla qubit is measured in the computational basis to obtain a binary outcome corresponding to the eigenvalue of its associated stabilizer ($+1$ or $-1$). Crucially, this projection extracts syndrome information without collapsing the logical state, thereby enabling non-demolition error detection.
\end{itemize}

\begin{figure}
    \centering
    \includegraphics[width=0.9\linewidth]{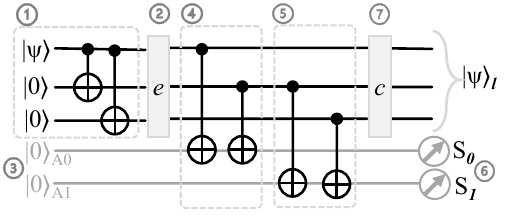}
    \caption{Demonstrating the quantum circuit that implements a $3-qubit$ repetition code. The circuit comprises the following components:
    (1) The state preparation circuit.
    (2) A single bit-flip error that may occur on any of the 3 qubits. 
    (3) Two ancilla qubits that are initialized to the state, $\ket{0}$ are employed for parity checking purposes.
    (4) The first stabilizer circuit, $Z \otimes Z \otimes I$, is responsible for measuring the parity between the first two qubits.
    (5) The second stabilizer circuit, $I \otimes Z \otimes Z$, measures the parity between the last two qubits.
    (6) The ancilla qubits are utilized to obtain the syndrome bits, $S_0, S_1$, which enable the detection and deduction of errors. 
    (7) To rectify the errors, a correction operator, $c$, comprising a sequence of self-inverse Pauli-gates, is applied to the qubit that requires correction. \cite{dummies}
    }
    \label{fig:3qec_rep_code}
\end{figure}

\begin{svgraybox}
\textbf{Case Study 1.4.1: The 3-Qubit Repetition Code \cite{dummies}: }

The fundamental ideas above are best illustrated through the 3-qubit repetition code, which protects against single-qubit bit-flip (Pauli-$X$) errors as seen in Fig. \ref{fig:3qec_rep_code}. The logical qubit $|\psi\rangle = \alpha|0\rangle + \beta|1\rangle$ is encoded into a three-qubit entangled state $|\psi\rangle_L = \alpha|000\rangle + \beta|111\rangle$ via a cascade of CNOT gates, where the original qubit acts as the control and two ancillary physical qubits serve as the targets.
After encoding, the stabilizer layer implements two parity-check operations:
\begin{itemize}
    \item $Z \otimes Z \otimes I$, which checks the parity between the first and second qubit.
    \item $I \otimes Z \otimes Z$, which checks the parity between the second and third qubit.
\end{itemize}
These stabilizers are implemented using two ancilla qubits, initialized in the state $|0\rangle$, which interact with the data qubits via controlled-NOT operations. The resulting ancilla states are measured to yield syndrome bits $S_0$ and $S_1$. The entire circuit layout is shown in Fig.~3, where each functional component---encoding, error insertion, stabilizer subcircuits, ancilla readout, and correction---is modularized.
This design enables the detection of single bit-flip errors on any of the three physical qubits. A lookup table (Table \ref{tab:qec_3qbit_syn}) associates each syndrome pattern $(S_0, S_1)$ with the location of the error, allowing an appropriate Pauli-$X$ correction (to be addressed in the decoding stage). Importantly, because stabilizer measurements are performed via ancilla projections, the encoded state remains coherent and unmeasured.
\end{svgraybox}
\begin{table}[t]
\caption{Detection, deduction, and correction of errors with respect to the syndrome measurements \cite{dummies}.}
\begin{center}
\begin{tabular}{||c | c | c | c | c||} 
\hline
 \multicolumn{2}{|c|}{\textbf{Detection}} & \multicolumn{2}{|c|}{\textbf{Deduction}} & \multicolumn{1}{|c|}{\textbf{Correction}}\\
 \hline
 \hline
 $S_0$ & $S_1$ & Error Location & Erroneous State & Correction Operator \\ [0.5ex] 
 \hline\hline
 $+1$ & $+1$ & No error & $\ket{000}$ & $III$ \\ 
 \hline
 $-1$ & $+1$ & Qubit 1 & $\ket{100}$ & $XII$ \\
 \hline
 $-1$ & $-1$ & Qubit 2 & $\ket{010}$  & $IXI$ \\
 \hline
 $+1$ & $-1$ & Qubit 3 & $\ket{001}$ & $IIX$ \\
 \hline
\end{tabular}
\label{tab:qec_3qbit_syn}
\end{center}
\end{table}

While the 3-qubit repetition code serves as a pedagogical example, the architectural template generalizes naturally to more sophisticated QECCs. In a generic \([[n, k, d]]\) stabilizer code, a logical register of $k$ qubits is encoded into a block of $n$ physical qubits. A set of $m = n - k$ stabilizer generators $\{g_1, g_2, \dots, g_m\}$ is employed to define the codespace and detect errors.
\begin{figure}[t]
    \centering
    \includegraphics[width=0.9\linewidth]{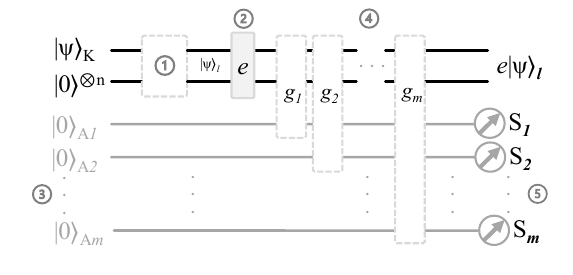}
    \vspace{-4mm}
    \caption{Illustration of a generalized stabilizer circuit that encodes $k$ physical qubits with $n$ logical qubits and employs $m$ stabilizers to detect potential errors in the logical codeword, $\ket{\psi}_l$, where $m=n-k$. The circuit comprises the following components: (1) The encoder circuit that encodes $k$ physical qubits with $n$ logical qubits; (2) A bit-flip or phase-flip or both errors that may occur on one or more qubits within the logical codeword, $\ket{\psi}_l$; (3) $m$ ancilla qubits that are initialized to $\ket{0}$ to facilitate the projection of the measurements of the stabilizer generators; (4) $m$ stabilizer generators, designated as $g_1, g_2, \dots, g_m$, are employed; and (5) Each stabilizer $g_i$ projects its measurements onto the ancilla qubit $\ket{0}_{Ai}$. The ancilla qubits are subsequently measured as syndrome measurements, $S_i$. Based on these syndrome values, error deduction and correction mechanisms are executed. It is essential to highlight that the provided diagram primarily illustrates a generalized stabilizer circuit designed for detecting errors. This representation does not encompass the execution of correction operators; consequently, the final basis state persists as the erroneous state, $e\ket{\psi}_l$ \cite{dummies}. }

    \vspace{-5mm}
    \label{fig:stab_form}
\end{figure}
The corresponding circuit-level realization, illustrated in Fig. \ref{fig:stab_form}, consists of the following components:
A \textbf{Unitary encoder} that maps $k$ logical qubits into $n$ entangled physical qubits forming the initial codeword $|\psi\rangle_L$;
\textbf{Ancilla qubits} $\{|0\rangle_{A_1}, \dots, |0\rangle_{A_m}\}$, each dedicated to measuring one stabilizer generator $g_i$;
 \textbf{Stabilizer measurement circuits}, wherein each $g_i$ is decomposed into a sequence of Pauli operations acting conditionally on the ancilla qubit. The interaction between data and ancilla qubits is governed by entangling gates, and measurement of the ancilla yields the syndrome bit $S_i$; and
the \textbf{Final logical state}, now possibly afflicted by correctable errors, is routed to downstream modules for decoding and correction.
This architectural abstraction underlies modern quantum codes, including surface codes, Bacon-Shor codes, and color codes, all of which instantiate stabilizer checks through structured ancilla-mediated circuits.

\subsection{QEC Codes}

Beyond the general stabilizer formalism, quantum error correction codes can be broadly categorized based on their structural principles, encoding strategies, and implementation constraints. In what follows, we outline four key families of QEC codes that have played pivotal roles in the development of fault-tolerant quantum architectures: concatenated codes, topological codes, subsystem codes, and bosonic codes.
\begin{itemize} [label={--}]
    \item \textbf{Concatenated Codes: } These codes recursively compose quantum codes to achieve exponential suppression of logical error rates \cite{knill1996concatenatedquantumcodes}\cite{gottesman1996pastingquantumcodes}\cite{Grassl_2009}. Given a base code \( \mathcal{C}_1 = [[n, k, d]] \) encoding \( k \) logical qubits into \( n \) physical qubits with code distance \( d \), one can encode each of the \( n \) physical qubits of \( \mathcal{C}_1 \) using another code \( \mathcal{C}_2 \), and so forth. The resultant code has an effective distance that grows multiplicatively across levels: for \( L \) levels of concatenation, the overall distance becomes \( d^L \), assuming identical base codes.
    Fault-tolerant gadgets for encoded gates, syndrome extraction, and error correction can be systematically constructed within this recursive structure. Classical control complexity and decoding remain tractable due to the tree-like modularity. A notable example is the concatenation of the \([[7,1,3]]\) Steane code \cite{Steane_1996}, which allows for transversal implementation of the Clifford group at each level. However, the total number of physical qubits scales as \( n^L \), leading to prohibitive overhead for large \( L \), especially in architectures constrained by spatial locality.

    \item \textbf{Topological Codes: } These encode logical information in the global degrees of freedom of a spatially extended lattice of qubits \cite{bombin2013introductiontopologicalquantumcodes}. These codes are defined via local stabilizer generators that act on a small, fixed number of neighboring physical qubits. Logical operators are non-local strings or loops that span the code lattice and commute with all stabilizers but are not part of the stabilizer group itself.
    Let \( \mathcal{S} \subset \mathcal{P}_n \) be the stabilizer group generated by local check operators. A topological code defines logical Pauli operators \( \bar{X}_i, \bar{Z}_i \in \mathcal{P}_n \) such that:
    \[
    \forall g \in \mathcal{S}, \quad [g, \bar{X}_i] = [g, \bar{Z}_i] = 0, \quad \text{and} \quad \bar{X}_i \bar{Z}_i = -\bar{Z}_i \bar{X}_i
    \]
    The most studied example is the \emph{surface code} \cite{Wootters1982}, defined on a 2D square lattice with two types of stabilizers: \emph{star operators} (X-type) and \emph{plaquette operators} (Z-type), each acting on four qubits. Logical qubits are encoded via nontrivial homology cycles across the lattice (in the toric code \cite{Kitaev_2003}) or via boundaries and defects (in the planar code \cite{freedman1998projectiveplaneplanarquantum}). The code distance \( d \) scales with the linear dimension of the lattice, and local syndrome measurements suffice for error detection.
    Topological codes exhibit high threshold error rates (e.g., \( \sim1\% \) for surface codes under depolarizing noise), and only require nearest-neighbor interactions, making them highly favorable for 2D hardware implementations \cite{2024}. However, realizing a universal gate set necessitates either lattice surgery, magic state distillation, or code switching, as not all logical operations can be implemented transversally.

    \item \textbf{Subsystem Codes: }In \emph{subsystem codes} \cite{aly2006subsystemcodes}, the codespace is partitioned as \( \mathcal{H}_{\text{code}} = \mathcal{H}_{\text{logical}} \otimes \mathcal{H}_{\text{gauge}} \), where \( \mathcal{H}_{\text{gauge}} \) is a gauge subsystem that carries no logical information \cite{Poulin_2005}. The stabilizer group \( \mathcal{S} \) is a subgroup of the (not necessarily abelian) gauge group \( \mathcal{G} \). Error correction is performed by measuring only a subset of \( \mathcal{G} \) (typically low-weight operators), using these to infer syndromes for the stabilizer subspace.
    An important example is the \emph{Bacon--Shor code} \cite{Bacon_2006}\cite{PhysRevA.52.R2493} \cite{bacon2003decoherencecontrolsymmetryquantum}, which protects against both bit-flip and phase-flip errors using two-dimensional repetition codes along orthogonal axes. The gauge operators are two-qubit Pauli operators along rows and columns, while logical operators span full rows or columns. Subsystem surface codes similarly introduce gauge degrees of freedom into surface codes to simplify syndrome extraction, particularly enabling the use of weight-2 checks instead of weight-4 or weight-6 stabilizers.
    The main advantage of subsystem codes lies in reduced measurement complexity and increased fault-tolerance in certain hardware settings. However, gauge freedom complicates decoding, and tradeoffs exist between gauge simplicity and code distance.

    \item \textbf{Bosonic Codes: }In these codes \cite{Grimsmo_2020}, the quantum information is encoded into continuous-variable modes rather than qubit-based systems. They exploit the infinite-dimensional Fock space of harmonic oscillators (e.g., photonic or phononic modes) and are particularly suited to superconducting cavity architectures or optical systems.
    The \emph{Gottesman--Kitaev--Preskill (GKP)} code \cite{https://doi.org/10.17169/refubium-45505}\cite{Conrad_2022} encodes a qubit into periodic grid-like superpositions of position eigenstates, enabling protection against small displacement errors in phase space. Logical operators are implemented as translations in quadrature space. The code stabilizers are displacement operators that define a lattice in the phase plane, and error correction involves measuring modular position and momentum.
    Other bosonic codes include \emph{cat codes} \cite{DODONOV1974597}\cite{Castanos1995CrystallizedCat}, which use superpositions of coherent states (e.g., \( |\alpha\rangle \pm |{-}\alpha\rangle \)), and \emph{binomial codes}\cite{Albert_2018}, which encode logical qubits using weighted superpositions of Fock states \cite{Rivera_2023} chosen to cancel specific error channels (e.g., photon loss). These codes allow for hardware-efficient error correction, particularly in platforms where bosonic modes exhibit longer coherence times than transmon qubits.
    Decoding bosonic codes involves a combination of continuous-variable measurement techniques and phase estimation, often integrated with feedback control. While they show promising error-correcting performance against dominant noise channels (e.g., amplitude damping), their performance under general noise remains an area of active research.
    
\end{itemize}
Among the various classes of quantum error correction codes, \emph{surface codes} have emerged as the leading candidate for scalable, fault-tolerant quantum computation \cite{2024}. Their construction on a 2D lattice of physical qubits with strictly local stabilizer measurements makes them well-suited for implementation in current quantum hardware architectures. Surface codes exhibit high error thresholds under realistic noise models and support flexible encoding schemes through boundaries and defects. Furthermore, their compatibility with efficient decoding algorithms and modular gate constructions, such as lattice surgery \cite{Chatterjee_2025}, renders them a practical foundation for large-scale FTQC. The following subsection explores the structure and operation of surface codes in detail.
\subsection{The Surface Code}
\begin{figure}
    \centering
    \includegraphics[width=1\linewidth]{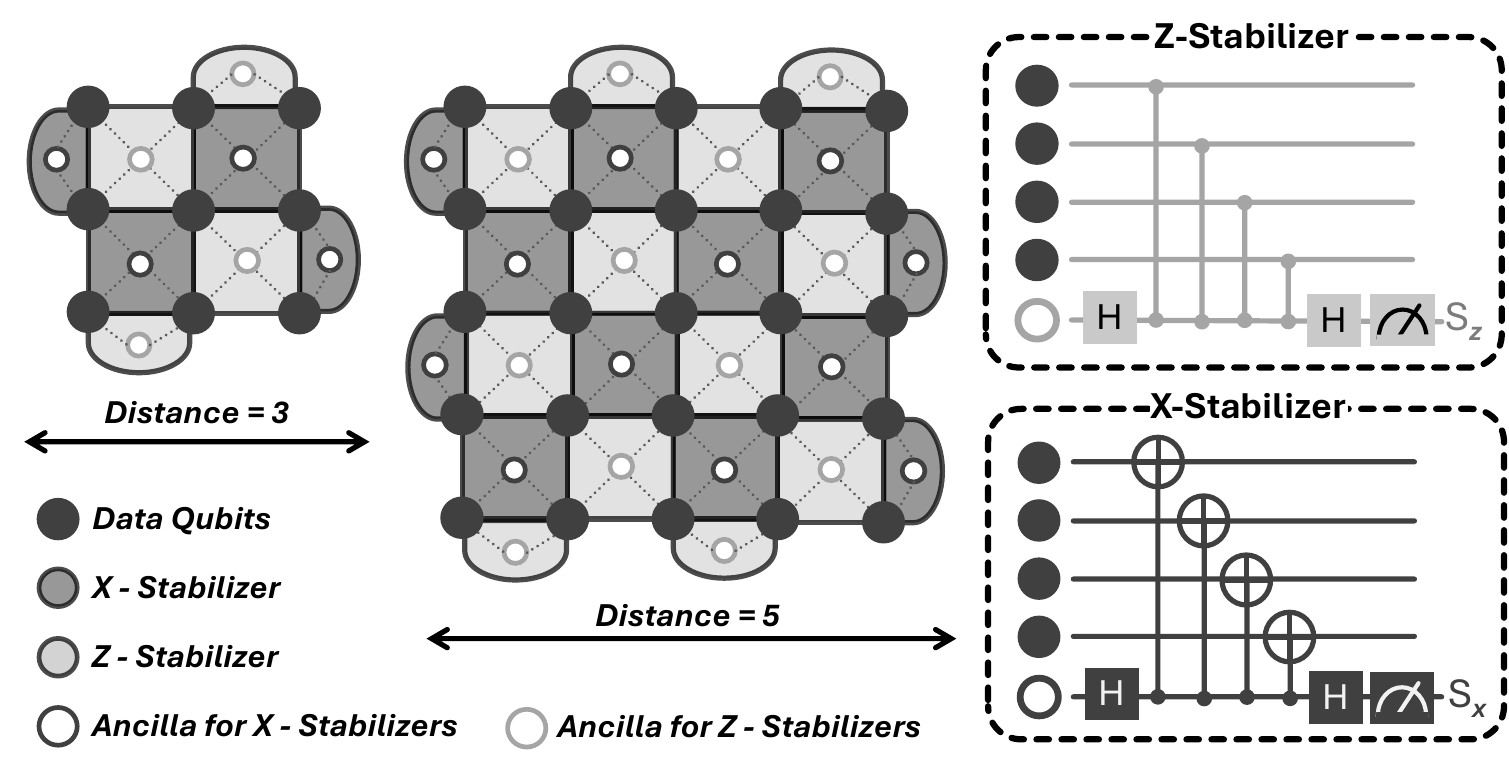}
    \caption{The diagrammatic representation of the structure of a rotated surface code.
    }
    \label{fig:surface_code}
\end{figure}
Surface codes (Fig. \ref{fig:surface_code}) constitute a class of topological stabilizer codes defined on a two-dimensional square lattice \cite{dummies}\cite{Wootters1982}. Each physical qubit is associated with an edge of the lattice, and the code's stabilizer group is generated by two sets of local operators: \emph{star operators} and \emph{plaquette operators}. For a given vertex \( s \), the corresponding star operator is defined as
\[
A_s = \prod_{i \in \text{star}(s)} X_i,
\]
where the product ranges over the four edges incident to \( s \), and \( X_i \) denotes the Pauli-X operator acting on qubit \( i \). Similarly, for each face or plaquette \( p \), the plaquette operator is defined as
\[
B_p = \prod_{i \in \text{plaquette}(p)} Z_i,
\]
with the product taken over the four edges surrounding the plaquette and \( Z_i \) the Pauli-Z operator. The set of all such \( A_s \) and \( B_p \) defines an abelian subgroup \( \mathcal{S} \subset \mathcal{P}_n \), where \( \mathcal{P}_n \) is the \( n \)-qubit Pauli group. These stabilizers mutually commute and can be measured locally, as each acts on at most four qubits.
By design, the stabilizer structure imposes a gauge symmetry on the lattice: the valid codespace is the simultaneous \( +1 \)-eigenspace of all stabilizer generators. Errors anticommute with a subset of the generators and thus produce detectable syndrome changes without collapsing the encoded logical information. 
To enable local measurement, each stabilizer is assigned a dedicated ancilla qubit located at the center of the corresponding vertex or plaquette. These ancilla qubits interact with the surrounding data qubits through a fixed entangling circuit during syndrome extraction. The overall layout ensures that all operations required for stabilizer measurement involve only nearest-neighbor interactions, making the architecture highly compatible with planar hardware designs.
At the boundaries of the lattice, stabilizer generators become truncated due to the reduced number of neighboring edges. This leads to the definition of two boundary types: \emph{smooth boundaries}, which terminate \( X \)-type stabilizers, and \emph{rough boundaries}, which terminate \( Z \)-type stabilizers. These boundary conditions not only affect the stabilizer structure but also play a crucial role in the definition of logical operators, as they determine the permissible endpoints of nontrivial Pauli operator strings.
\begin{figure}
    \centering
    \includegraphics[width=0.9\linewidth]{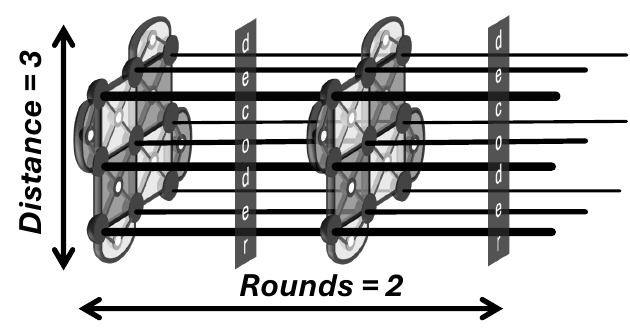}
    \caption{A diagrammatic representation of running a surface code of distance 3 for 2 \emph{rounds}.
    }
    \label{fig:decoding_rounds}
\end{figure}
\subsubsection*{Logical Qubits and Code Distance}
Logical qubits in a surface code are encoded by exploiting the global topological features of the lattice. In the planar surface code, logical operators correspond to nontrivial paths of Pauli operators that connect specific types of boundaries. Two types of boundaries are defined on the lattice: \emph{rough boundaries}, which terminate \( Z \)-type plaquette stabilizers, and \emph{smooth boundaries}, which terminate \( X \)-type star stabilizers.
A logical \( \overline{Z} \) operator is implemented by a chain of Pauli-\( Z \) operators acting on qubits along a path that connects two rough boundaries. Similarly, a logical \( \overline{X} \) operator is a chain of Pauli-\( X \) operators that connects two smooth boundaries. These logical operators commute with all stabilizers but are not themselves part of the stabilizer group, and they satisfy the requisite anticommutation relation \( \overline{X} \, \overline{Z} = - \overline{Z} \, \overline{X} \).
The \emph{code distance} \( d \) is defined as the minimum weight (i.e., the number of qubits acted upon) of any Pauli operator that implements a nontrivial logical operation. For a square lattice of linear size \( d \), the shortest logical operator spans \( d \) physical qubits, and the code can thus correct up to \( \lfloor (d - 1)/2 \rfloor \) arbitrary single-qubit errors. 
Surface codes perform stabilizer measurements in repeated \textit{rounds} to extract syndromes over time. Multiple rounds—typically \( d \)—are needed to distinguish data errors from measurement noise. These rounds define the temporal dimension of decoding graphs, enabling identification of error chains via algorithms such as minimum-weight perfect matching on space-time syndrome data (Fig. \ref{fig:decoding_rounds}).
The number of logical qubits encodable depends on the geometry of the patch: for a planar code with an appropriate boundary configuration, a single logical qubit is typically encoded per patch.

\subsubsection*{Syndrome Extraction}
Syndrome extraction in surface codes is performed by measuring the stabilizer generators \( A_s \) and \( B_p \) repeatedly over time using dedicated ancilla qubits. Each ancilla qubit is associated with a specific stabilizer and interacts with the data qubits on the corresponding star or plaquette via a fixed sequence of entangling gates.
To extract the eigenvalue of a star operator \( A_s \), the corresponding ancilla qubit is initialized in the \( |+\rangle \) state, entangled with the neighboring data qubits through a series of $CNOT$ gates (with the ancilla as control), and then measured in the \( X \)-basis. A similar procedure applies for plaquette operators \( B_p \), using ancilla initialization in the \( |0\rangle \) state, $CNOT$ gates with the ancilla as target, and measurement in the \( Z \)-basis.
A nontrivial measurement outcome (i.e., \( -1 \)) indicates that an error has occurred on one or more of the neighboring data qubits that anticommute with the corresponding stabilizer. Since errors affect the measured eigenvalue rather than the data state itself, the extraction process is nondestructive. Repeating syndrome measurements over multiple cycles enables temporal redundancy, which is essential for identifying and correcting time-correlated or measurement-induced errors.
To achieve full fault tolerance, syndrome extraction must be repeated over multiple time steps. For a surface code of distance \( d \), it is necessary to perform \( d \) rounds of syndrome measurements in order to reliably detect and correct time-correlated errors, including faults arising during ancilla preparation and measurement. This repetition effectively extends the decoding problem into three dimensions, where errors are represented as strings in a \( 2+1 \) dimensional space-time volume. Decoders must account for both spatial and temporal correlations, ensuring that physical and measurement errors can be simultaneously corrected without logical failure.

\subsubsection*{Decoding in Surface Codes}
The decoding process \cite{wu2024legoqecdecodingarchitecture}\cite{wu2023fusionblossomfastmwpm} operates on the space-time syndrome history generated by repeated rounds of stabilizer measurements, as described in the previous section, involving the identification of the most likely configuration of physical errors that could have produced the observed stabilizer syndrome. Since each stabilizer generator either commutes or anticommutes with a single-qubit Pauli error, a local error manifests as a violation (i.e., a \(-1\) measurement outcome) of one or more stabilizer checks. These violations, also called \emph{syndrome defects}, are detected at the endpoints of error chains.
In this framework, errors are naturally represented as \emph{strings} of Pauli operators acting on connected paths of qubits along the lattice \cite{wang2023transformerqecquantumerrorcorrection}. A single Pauli error introduces a pair of syndrome defects at its adjacent stabilizers. More generally, a chain of single-qubit errors will produce syndrome defects at the endpoints of the string, while the internal stabilizers along the path remain satisfied due to cancellation of anticommuting effects. Logical errors correspond to nontrivial homologically inequivalent strings that span the code and connect distinct boundary types, potentially changing the encoded logical state if undetected.
The task of the decoder is to infer the most probable set of error strings that match the observed syndrome data, ideally correcting all errors without inducing a logical fault. Under the assumption of independent and identically distributed noise, the decoding problem reduces to a minimum-weight matching problem on a graph where nodes represent syndrome defects and edge weights reflect error probabilities \cite{wu2023fusionblossomfastmwpm}.

\subsubsection*{Logical Operations and Code Variants}
Surface codes support fault-tolerant implementation of logical operations through topological manipulations of the encoded qubits. While transversal gates are severely limited by the Eastin–Knill theorem \cite{Eastin_2009}, the surface code architecture admits alternative schemes such as \emph{lattice surgery} \cite{Chatterjee_2025} and \emph{code deformation} \cite{Dua_2024}. In lattice surgery, logical operations like $CNOT$ are realized by dynamically merging and splitting code patches via stabilizer reconfiguration along shared boundaries. Clifford group operations can thus be executed in a fault-tolerant manner using purely local interactions and syndrome measurements \cite{Litinski_2019}.
\emph{Rotated surface codes} \cite{Bombin_2007} \cite{anderson2011homologicalstabilizercodes} modify the lattice geometry to achieve the same code distance with fewer physical qubits. The \emph{$XZZX$ code} \cite{Wen_2003} \cite{Bonilla_Ataides_2021} \cite{Kovalev_2011} alternates the type of Pauli operators in stabilizers across the lattice, making it particularly effective under biased noise conditions. More advanced schemes based on defects \cite{Nagayama_2017}, holes \cite{Brown_2017}, or twist operators \cite{Litinski_2018} allow the encoding of multiple logical qubits within a single patch and support braiding-based gate constructions. These variants retain the essential features of locality and high threshold while offering different trade-offs in encoding density and decoding complexity.

\subsection{Lattice Surgery}
\begin{figure}
    \centering
    \includegraphics[width=1\linewidth]{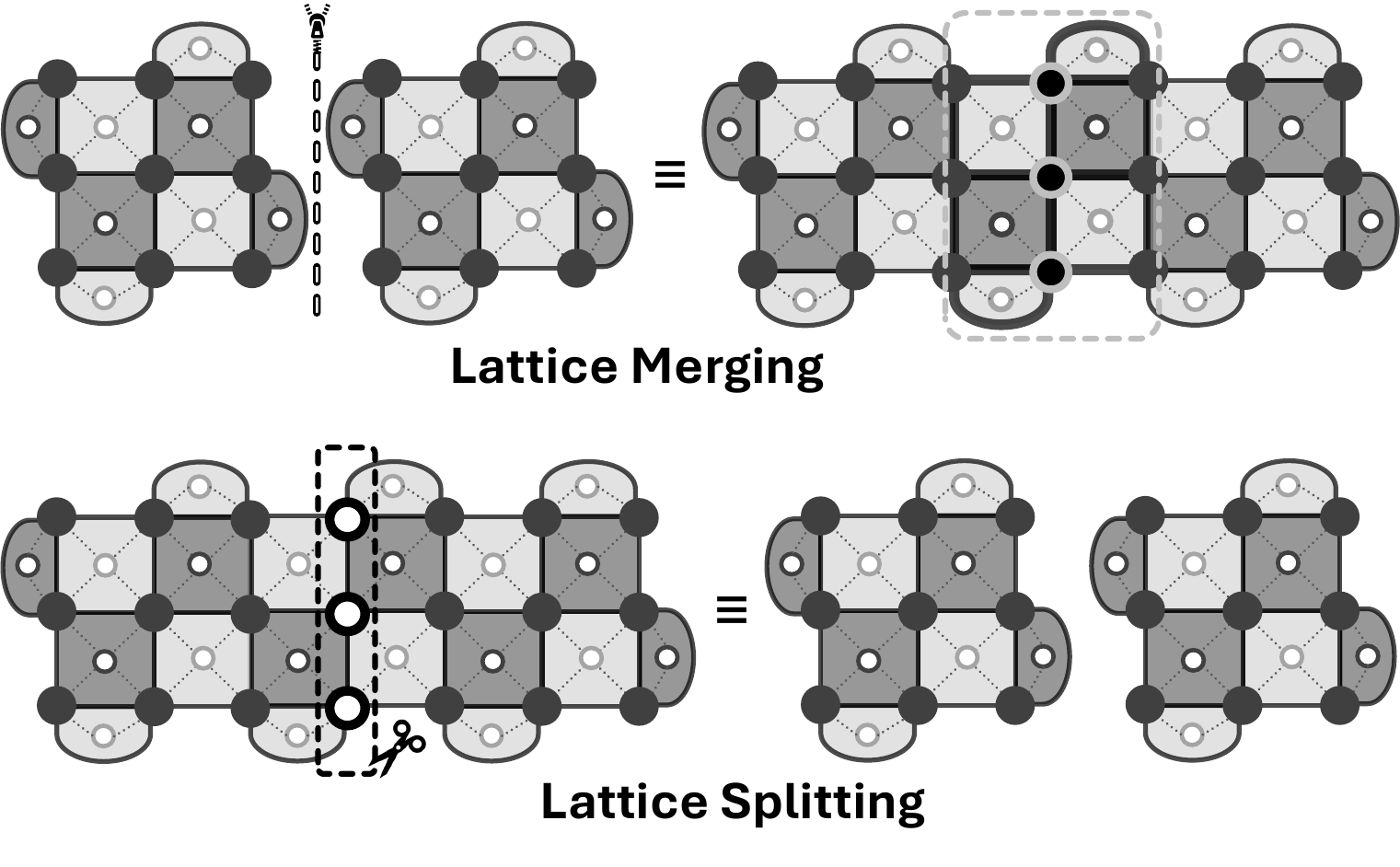}
    \caption{Diagrammatic representation of the merging and splitting of the lattices in lattice surgery.
    }
    \label{fig:lattice_surgery}
\end{figure}

Surface codes operate under the constraint of a two-dimensional nearest-neighbor (2DNN) interaction topology, wherein physical qubits are arranged in a planar grid and interact exclusively with their adjacent neighbors \cite{Litinski_2018}. This constraint aligns naturally with the physical architecture of many leading quantum hardware platforms, particularly superconducting qubits, and enables fault-tolerant quantum computation via local stabilizer measurements.
In the surface code framework, each logical qubit is encoded across a two-dimensional lattice of physical qubits. Executing a quantum algorithm thus entails assigning a distinct surface to every logical qubit, each supporting stabilizer measurement cycles for error detection and correction. As the logical width of a quantum algorithm increases, so too does the spatial demand for separate, concurrently active logical surfaces.
While certain logical gates—such as \( X \), \( Z \), or $H$—can be implemented transversally within a single logical surface, two-qubit gates such as CNOT generally necessitate interactions between distinct logical patches. In a strict 2DNN architecture, the transversal application of two-qubit gates across distant surfaces violates locality constraints and imposes significant overheads in connectivity and error suppression.
To address this, \emph{lattice surgery} provides an alternative paradigm that facilitates logical operations via local boundary manipulations rather than global transversal gates. The key idea is to dynamically modify the topology of logical patches through operations known as \textit{merging} and \textit{splitting}. These operations, when executed along appropriate stabilizer boundaries (e.g., $X$-type or $Z$-type), enable the implementation of entangling gates and joint measurements entirely within the 2DNN model \cite{Horsman_2012}\cite{Chatterjee_2025}.

\subsubsection*{Merge and Split Operations}
The fundamental operations in lattice surgery are defined as follows (Fig. \ref{fig:lattice_surgery}):
\begin{itemize}[label={--}]
    \item \textbf{Merge:} Two adjacent logical surfaces are merged along a common boundary to perform a joint stabilizer measurement. For instance, a merge along the $X$-boundary enables the measurement of the joint \( X \otimes X \) operator. This is achieved by modifying the stabilizer configuration across the boundary—specifically, by introducing new ancillary qubits between the patches and redefining the stabilizers to include these ancillae and boundary qubits. The outcome of the merge corresponds to a parity measurement between the two logical qubits, enabling fault-tolerant implementation of certain Clifford operations and state preparations.

    \item \textbf{Split:} The inverse of merging, splitting separates a previously merged surface into two independent logical patches. This is effected by measuring the ancillary qubits between the two logical regions in a basis (typically \( Z \)) that disentangles their stabilizer structure. The measurement outcomes are used to update the Pauli frame accordingly. After splitting, each surface resumes independent stabilizer measurement cycles, thus restoring individual logical qubits.
\end{itemize}
Merge and split operations can be performed along either $X$- or $Z$-type boundaries, depending on the desired logical operation. The process is entirely measurement-based, and all operations are confined to local stabilizer redefinitions and ancilla-mediated interactions. This locality is preserved at every step, thereby maintaining full compatibility with the 2DNN constraint and ensuring fault tolerance throughout.

\subsubsection*{Implications for Modular Architectures}
Lattice surgery underpins many modern proposals for modular, patch-based quantum architectures \cite{Litinski_2019} \cite{Litinski_2018}. In particular, the ability to entangle, disentangle, and measure logical qubits through controlled reshaping of code patches facilitates the design of dedicated \textit{data blocks} for computation and \textit{distillation blocks} for magic state preparation. Communication between these blocks is mediated by lattice surgery operations, enabling scalable and efficient resource management without necessitating qubit transport or long-range interactions.

%% file: author/section2.tex
\section{The QEC Design Flow}
\label{sec2}
This subsection outlines the design flow of a quantum error correction (QEC) protocol, tracing the transformation from logical circuit synthesis through Pauli product rotations, Clifford gate pruning, and Clifford+$T$ decomposition, culminating in magic state distillation and physical realisation. The flow captures key transformations enabling FTQC.

\subsection{Logical Circuit Synthesis}

Logical circuit synthesis constitutes the first transformation in the QEC design flow. It bridges the gap between a high-level algorithmic description of quantum computation and a low-level, gate-based representation suitable for encoding under fault-tolerant quantum error correction schemes. The core objective is to represent the entire computation using a universal set of operations that are compatible with fault-tolerant implementation, particularly under stabilizer codes such as the surface code.

\subsubsection*{Pauli Product Rotation Formalism}

Rather than immediately targeting a conventional gate set such as Clifford+$T$, recent advances propose an algebraically structured representation of quantum circuits using \emph{Pauli product rotations} \cite{Litinski_2019}. Each quantum gate is expressed as a unitary of the form
\[
P_\varphi = \exp(-i \varphi P),
\]
where \(P\) is a tensor product of Pauli operators (\(I, X, Y, Z\)) acting on one or more qubits, and \(\varphi\) is a rotation angle.
This representation naturally subsumes all standard single- and multi-qubit gates:
\begin{align*}
S &= Z_{\pi/4}, \quad T = Z_{\pi/8}, \\
H &= Z_{\pi/4} X_{\pi/4} Z_{\pi/4}, \\
\mathrm{CNOT} &= (Z \otimes X)_{\pi/4} \cdot (I \otimes X)_{-\pi/4} \cdot (Z \otimes I)_{-\pi/4}.
\end{align*}
More generally, a controlled-\(P\) gate, \(C(P_1,P_2)\) can be written as:
\[
C(P_1, P_2) = (P_1 \otimes P_2)_{\pi/4} \cdot (I \otimes P_2)_{-\pi/4} \cdot (P_1 \otimes I)_{-\pi/4}.
\]
This algebraic formalism unifies gate synthesis, gate commutation, and cancellation within a single operator-theoretic framework that is particularly well-suited for fault-tolerant execution, aligning with surface-code lattice surgery, which naturally implements such rotations via multi-qubit Pauli product measurements \cite{Lai_2022}\cite{coecke2018zxrules2qubitcliffordtquantum}\cite{Litinski_2019}.

\subsubsection*{Clifford Elimination and Gate Commutation}
Clifford gates, by virtue of mapping Pauli operators to Pauli operators under conjugation, can be commuted to the end of the circuit \cite{Litinski_2019}\cite{Pllaha_2021}\cite{patel2003efficientsynthesislinearreversible}. This is a key transformation step, as Clifford operations can typically be implemented transversally or classically tracked, whereas non-Clifford gates (e.g., \(T\)) must be realized through magic state injection and distillation \cite{Litinski_2019_msd}.
For any Clifford operator \(C\) and Pauli product rotation \(P_\varphi\), one has:
\[
C P_\varphi C^\dagger = P'_\varphi,
\]
where \(P'\) is again a Pauli product. Hence, any quantum circuit \(U\) can be transformed into a canonical form:
\[
U = C_f \cdot \left(\prod_j P^{(j)}_{\pi/8}\right) \cdot \left(\prod_k P^{(k)}_{\pi/4}\right),
\]
where \(C_f\) is the final Clifford layer. 
This transformation has two practical consequences--
(1) All \emph{Clifford gates} are deferred to the end of the quantum circuit and eventually absorbed into the measurement bases by the following rules: 
\[
P_{\pi/4} \cdot P'_\varphi =
\begin{cases}
P'_\varphi P_{\pi/4}, & \text{if } [P, P'] = 0, \\
(iPP')_\varphi P_{\pi/4}, & \text{if } \{P, P'\} = 0.
\end{cases}
\]
and 
(2) the transformed circuit consists exclusively of \emph{Pauli product rotations}, organized according to their rotation angles. All non-Clifford operations are executed via magic-state-assisted Pauli product measurements, while the residual Clifford action is accounted for through a classical update of the final measurement bases.

\subsubsection*{Clifford+$T$ Formalism}

After commutation, the resulting synthesized circuit adopts a structure tailored for QEC-aware transpilation:
\begin{itemize}[label={--}]
    \item A sequence of \(\pi/8\) rotations, i.e., \(T\) gates, each expressible as \(P_{\pi/8}\),
    \item A sequence of \(\pi/4\) Clifford rotations, i.e., \(P_{\pi/4}\),
    \item A terminal layer of Pauli product measurements, where the final Clifford gates have been absorbed.
\end{itemize}
This \emph{canonical Clifford+$T$ form} \cite{dawson2005solovaykitaevalgorithm} \cite{ross2016optimalancillafreecliffordtapproximation} \cite{Bravyi_2005} ensures that all operations beyond the Clifford group are explicitly isolated and ready for translation into fault-tolerant primitives. Each \(T\) gate is implemented via ancillary magic states, while all Clifford operations can be transversally encoded or classically managed.

Therefore, the original quantum circuit, when formulated via the Pauli product rotations and Clifford commutation, yields a representation of the form:
\[
\text{Input circuit} \quad \Rightarrow \quad \{\pi/8\text{-rotations}\} + \{\pi/4\text{-rotations}\} + \{\text{Pauli measurements}\}
\]

\subsection{Magic State Distillation}
The Clifford+$T$ decomposition forms a universal gate set for quantum computation, wherein Clifford operations can typically be implemented fault-tolerantly and transversally under stabilizer codes. However, the non-Clifford \(T\) gate presents a fundamental challenge: due to the Eastin-Knill theorem \cite{Eastin_2009}, no quantum error-correcting code that admits universal quantum computation allows all gates, including \(T\), to be implemented transversally. As a result, the standard method for implementing the \(T\) gate in fault-tolerant architectures is via \emph{magic state injection}.
A \emph{magic state} is a specially prepared non-stabilizer ancilla qubit in the state
\[
|m\rangle = \frac{1}{\sqrt{2}}(|0\rangle + e^{i\pi/4} |1\rangle),
\]
which, when consumed in an ancilla-assisted measurement circuit, allows for the indirect application of the \(T\) gate on a data qubit, followed by a classically controlled Clifford correction. However, such magic states, when prepared at the physical level, are highly susceptible to noise and decoherence, rendering them unsuitable for direct use. Their error rates typically exceed the fault-tolerance threshold of the code and must be reduced via a process known as \emph{magic state distillation}  \cite{Campbell_2017} \cite{Howard_2014} \cite{knill2004faulttolerantpostselectedquantumcomputation} \cite{Bravyi_2012}\cite{Bravyi_2005_cliff} \cite{Bartlett2014}.
\subsubsection*{Distillation Protocols}
Magic state distillation employs quantum codes that admit transversal \(T\) gates to recursively suppress noise in faulty ancillae. Classical protocols, such as the Bravyi-Kitaev 15-to-1 scheme based on the \([15,1,3]\) punctured Reed-Muller code \cite{Litinski_2019_msd}\cite{Litinski_2019}\cite{guruswami2017efficientlylistdecodablepuncturedreedmuller}, take \(15\) noisy magic states as input and output a single, higher-fidelity magic state. The procedure involves the following steps:
\begin{enumerate}
    \item Encoding the input states into the logical subspace of a distillation code.
    \item Applying a transversal \(T\) gate across all qubits.
    \item Decoding and measuring stabilizers to detect the presence of uncorrectable errors.
    \item Postselecting on syndrome outcomes to accept or reject the output state.
\end{enumerate}
This protocol reduces the error probability from \(p\) to approximately \(35p^3\) under an independent depolarizing noise model \cite{Litinski_2019_msd}\cite{Bravyi_2012}. More generally, distillation protocols are characterized by triples \((n,k,d)\), indicating \(n\) input states, \(k\) outputs, and distance \(d\), often realized through triorthogonal matrices and CSS constructions. Following the magic state distillation process, the distilled states are placed and routed on the data patches on the 2D lattice of qubits, optimizing for the space-time cost \cite{Litinski_2019}.

\subsection{Decoding}
The decoding stage constitutes the classical inference process responsible for identifying and correcting quantum errors based on stabilizer measurement outcomes \cite{wu2024legoqecdecodingarchitecture}\cite{wu2023fusionblossomfastmwpm}\cite{wang2023transformerqecquantumerrorcorrection}. Given that direct measurement of data qubits is prohibited in fault-tolerant protocols to preserve quantum coherence, QEC architectures employ ancillary measurements of stabilizer generators to indirectly reveal the presence and type of errors. The decoding problem, then, is to map these syndrome outcomes to appropriate correction operations, with the goal of recovering the original logical state without introducing further logical errors.
Mathematically, decoding can be viewed as a function
\[
D: \{0,1\}^m \to \mathcal{P}_n,
\]
where \( \{0,1\}^m \) is the binary syndrome space generated by \( m \) stabilizer measurements, and \( \mathcal{P}_n \) is the \( n \)-qubit Pauli group, from which a recovery operator is chosen. A decoder is thus a classical algorithm that infers an error coset representative from a syndrome string, ideally identifying the most probable equivalence class of errors consistent with the measured syndrome.
The decoding process is intimately tied to the error model and the structure of the quantum code. Under a stochastic Pauli noise model, each error \( E \in \mathcal{P}_n \) commutes or anticommutes with each stabilizer generator \( S_i \), and the syndrome bit \( s_i \) is defined as:
\[
s_i = 
\begin{cases}
0 & \text{if } [E, S_i] = 0, \\
1 & \text{if } \{E, S_i\} = 0.
\end{cases}
\]
Since multiple errors can yield the same syndrome---forming a coset of the stabilizer group---the decoding task is fundamentally probabilistic. A decoder aims to return a correction operator \( C \) such that \( CE \in \mathcal{S} \cdot \mathcal{L}_0 \), where \( \mathcal{S} \) is the stabilizer group and \( \mathcal{L}_0 \) is the identity logical operator. Any deviation from this results in a logical error.

\subsubsection*{Code-Specific Decoding Strategies}
The decoding algorithm employed depends heavily on the structure of the underlying quantum code.

\begin{itemize}[label={--}]
    \item \textbf{Repetition Code:} In the \( n \)-qubit bit-flip repetition code, the syndrome reveals the parity of adjacent qubits \cite{PhysRevA.111.012419}\cite{Cirelson1978}. Decoding is performed by majority voting: the value shared by the majority of physical qubits is taken as the logical value.

    \item \textbf{Surface Code:} The most widely used code in practical architectures, surface codes define syndrome extraction in terms of vertex ($X$-type) and plaquette ($Z$-type) stabilizers on a two-dimensional lattice. Decoding is performed by matching the syndrome defects (syndrome bits equal to 1) using the \emph{minimum-weight perfect matching} (MWPM) algorithm \cite{Wu_2025} \cite{wu2023fusionblossomfastmwpm}. This algorithm pairs detection events via a graph matching procedure that minimizes total edge weight, which corresponds to error likelihoods.

    \item \textbf{Color Code:} These codes, defined on trivalent lattices with three-colorable faces, present additional decoding challenges due to higher-weight stabilizers and overlapping supports. Decoders for color codes often rely on generalizations of matching techniques or employ message passing and belief propagation \cite{Lee_2025}.

    \item \textbf{Subsystem Codes:} These codes introduce gauge degrees of freedom, further complicating decoding. In such cases, decoding typically proceeds in two stages: gauge fixing and then syndrome interpretation on the resulting stabilizer code \cite{sarvepalli2008encodingsubsystemcodes}.
\end{itemize}

\subsubsection*{Performance Criteria and Noise Models}
The performance of a quantum decoder is evaluated along several critical dimensions, including accuracy, latency, scalability, and fault tolerance \cite{wang2023transformerqecquantumerrorcorrection} \cite{wu2023fusionblossomfastmwpm}. Accuracy refers to the probability that the decoder returns a correction operator lying in the same coset as the true error, thereby preserving the logical state. Latency measures the computational delay incurred during decoding, which is particularly important for real-time feedback in fault-tolerant quantum computing. Scalability assesses the algorithm's complexity and feasibility as the code distance increases, especially relevant for large-scale codes such as surface or LDPC codes. Fault tolerance encompasses the decoder's robustness against ancillary imperfections, such as measurement errors or correlated faults, ensuring that the decoding process does not itself become a source of logical failure. These performance metrics are inherently dependent on the assumed noise model, which may include depolarizing noise (uniform Pauli errors), biased noise (e.g., dephasing-dominated models), erasure noise (common in photonic architectures), and spatially or temporally correlated noise. Additionally, practical implementations must often incorporate measurement errors, necessitating syndrome extraction over multiple rounds and the deployment of temporal decoding strategies that track syndrome histories across time to distinguish data and measurement faults.

The decoding procedure must be adapted to the specific characteristics of the underlying noise model, as the structure and correlations in physical errors directly influence syndrome patterns and decoder performance. Common noise models include depolarizing noise, where all single-qubit Pauli errors occur with equal probability; biased noise, such as dephasing-dominated channels where one Pauli error (typically \( Z \)) is more prevalent; erasure noise, wherein qubit loss is detectable and localized—frequently encountered in photonic systems; and correlated noise, arising from crosstalk, leakage, or environmental coupling, which induces multi-qubit error events. Additionally, measurement errors compromise the reliability of stabilizer readouts, necessitating repeated syndrome extraction across multiple error-correction cycles. To address this, modern decoders often employ \emph{temporal decoding}, wherein sequences of syndromes are jointly analyzed to differentiate between data and measurement faults. In topological codes such as the surface code, this results in a space-time decoding framework, typically modeled as a three-dimensional syndrome graph on which algorithms like minimum-weight perfect matching (MWPM) are applied to infer the most likely error trajectories through both space and time.

\subsubsection*{Classical Decoding Algorithms}

Several classical decoding schemes are employed in practice:

\begin{itemize}[label={--}]
    \item \textbf{Lookup Table Decoders:} These decoders enumerate all possible syndromes and their corresponding optimal corrections in a precomputed table. For each observed syndrome, the decoder simply returns the associated correction operator. This method guarantees exact decoding under the assumed noise model and is computationally trivial at runtime \cite{das2021lilliputlightweightlowlatencylookuptable}. However, the space complexity grows exponentially with the number of stabilizers, making this approach infeasible beyond small codes such as the 3-qubit repetition code or the 5-qubit perfect code. Lookup table decoders are primarily useful for benchmarking purposes and for pedagogical illustration.

    \item \textbf{Minimum-Weight Perfect Matching (MWPM):} Widely employed in decoding surface codes, MWPM \cite{wu2023fusionblossomfastmwpm}\cite{wu2024legoqecdecodingarchitecture}\cite{Wu_2025} operates on a graph constructed from syndrome data, where each node represents a detection event (typically a non-trivial stabilizer measurement outcome) and weighted edges correspond to the likelihood or cost of error paths connecting them. Under the assumption of independent errors, the decoder seeks a perfect matching that minimizes the total weight, i.e., the sum of edge costs. This is classically solved using Edmonds' blossom algorithm in \( O(n^3) \) time for \( n \) nodes, though optimizations and parallel implementations exist. The error chains corresponding to the matched paths are applied as corrections. MWPM is effective under low noise rates and local stochastic noise models but does not naturally generalize to codes with non-planar structures or correlated noise.

    \item \textbf{Union-Find Decoders:} These decoders offer a computationally efficient alternative to MWPM, operating in nearly linear time. The core idea is to treat syndrome defects as seeds of growing clusters and perform disjoint-set operations to identify and merge connected components. When clusters overlap or reach a predefined radius, corrections are inferred by identifying minimal connecting paths between defects. The union-find data structure supports efficient merging and path compression, significantly reducing decoding latency \cite{Chan_2023}\cite{Griffiths_2024} \cite{val_uf}. While union-find decoders may not always match the accuracy of MWPM, especially near the code threshold, they scale better with lattice size and are promising candidates for hardware-level integration.

    \item \textbf{Belief Propagation (BP):} Particularly suited to quantum low-density parity-check (LDPC) codes, belief propagation is an iterative algorithm that operates on a Tanner graph representing stabilizer checks and physical qubits \cite{yao2024beliefpropagationdecodingquantum}\cite{kuo2024faulttolerantbeliefpropagationpractical}. Messages are passed between variable nodes (qubits) and check nodes (stabilizers), updating beliefs about the marginal probability of errors on each qubit conditioned on neighboring syndromes. The decoder converges to a configuration with maximum posterior probability under the assumption of a factorized prior. While BP is efficient for sparse graphs, it can fail to converge or yield suboptimal results in the presence of short cycles or strong correlations. Nevertheless, it provides a practical decoding strategy for large, sparse codes where exact inference is intractable.
\end{itemize}

Each decoding method makes specific assumptions about the noise model, code structure, and computational resources available. While optimal decoding (e.g., maximum likelihood decoding) is in general NP-hard \cite{wang2023transformerqecquantumerrorcorrection}\cite{wu2023fusionblossomfastmwpm}, these heuristic or approximate decoders offer practical trade-offs for real-world fault-tolerant quantum computing architectures.

%% file: author/section3.tex
\section{Design Automation and Optimization in QEC}
\label{sec3}

The design and implementation of large-scale quantum error-corrected computations demand rigorous optimization across multiple abstraction levels. In classical computing, Electronic Design Automation (EDA) \cite{eda1}\cite{eda2}\cite{eda3}\cite{Chen_2024} encompasses a suite of algorithmic tools and methodologies for the automated synthesis, placement, routing, and verification of digital circuits, whether targeting ASICs or FPGAs. Classical EDA frameworks are driven by a well-defined set of design objectives—primarily area, timing (latency), power consumption, and functional correctness—subject to physical constraints imposed by the technology node and fabrication process. Over several decades, EDA has matured into a highly structured and automated domain, with established toolchains and optimization strategies.
In quantum computing, particularly in the context of fault-tolerant quantum error correction (QEC), the design landscape is still nascent \cite{ghosh2025survivaloptimizedevolutionaryapproach}. However, the need for optimization is no less critical. Quantum resources are scarce, noisy, and expensive to operate; thus, optimizing their usage is essential for the viability of any large-scale quantum computation. The QEC design and compilation flow, although structurally different from its classical counterpart, inherits several optimization objectives. These can be broadly categorized as follows:

\begin{itemize}[label={--}]
    \item \textbf{Resource Minimization:} The total number of physical qubits required—including data, ancilla, and magic state qubits—must be minimized to conform to the limited quantum hardware available. This also includes reducing the demand for magic state distillation, which dominates the space-time overhead in fault-tolerant architectures.
    
    \item \textbf{Latency and Depth Reduction:} Minimizing logical circuit depth directly impacts the overall time for computation and reduces cumulative decoherence and operational error. In surface code-based realizations, this translates to minimizing the number of sequential lattice surgery steps or measurement rounds.
    
    \item \textbf{Fidelity Preservation under Formal Verification:} Employing quantum model checking or SMT‐based equivalence to ensure the optimized circuit maintains or improves logical‐state fidelity and introduces no new uncorrectable error paths.
\end{itemize}

These objectives are constrained by the fault-tolerance requirements of the quantum code: gates must respect transversal or code-compatible implementations, ancilla states must be prepared and verified without measurement collapse, and operations must preserve the logical subspace. Unlike classical logic synthesis, quantum design workflows must also manage non-Clifford gate costs, non-determinism in measurement-based logic (e.g., teleportation or lattice surgery), and hardware-induced connectivity restrictions. As a result, the optimization of QEC workloads poses novel algorithmic challenges that require specialized automation strategies.

\subsection{Reducing qubit overhead}

The robustness afforded by QEC comes at the cost of significant physical resource overhead, often orders of magnitude larger than the logical information being protected\cite{dummies}. This overhead manifests in multiple forms, including the spatial footprint of encoded data qubits, ancillary qubits required for stabilizer measurements, additional patches for routing and code deformation, and the redundancy needed to tolerate hardware defects or adversarial noise \cite{Litinski_2019}.
In surface code-based QEC architectures, which are widely regarded as a leading candidate for practical implementation due to their high logical error thresholds and local interactions, the qubit overhead scales approximately quadratically with the code distance. For a code of distance $d$, the number of data qubits is $\mathcal{O}(d^2)$, and the number of ancilla qubits is typically of the same order \cite{qprom}. Furthermore, to ensure reliable computation over large programs, the architecture must accommodate both spatial and temporal variations—such as fabrication-induced defects, cosmic ray-induced transient faults, and workload-induced congestion—which further increase the qubit requirement via dynamic deformation or redundancy provisioning \cite{shor1997faulttolerantquantumcomputation}.

As the field transitions from theoretical code design to practical realization, the challenge of qubit overhead has emerged as a critical bottleneck. This has motivated a growing body of research focused on automating overhead-aware QEC design. Recent research has explored the areas widely, from modifying the structure of the surface code by forming superstabilizers around fault regions \cite{Siegel_2023} to enabling localized code deformation and dynamic patch resizing to recover lost code distance while minimizing physical footprint \cite{yin2024surfdeformermitigatingdynamicdefects} as well as attempting to reduce the total qubit usage by reusing ancilla qubits across $X-$ and $Z-$type stabilizer measurements through temporally interleaved sub-rounds \cite{qprom}.

\begin{svgraybox}
\textbf{Case Study 3.1.1: Adaptive Surface Code \cite{Siegel_2023}}

The adaptive surface code framework introduces a dynamic method to preserve surface code functionality in the presence of temporary or permanent defective qubits by replacing stabilizers that intersect the defect region with superstabilizers. This enables the code to continue operating without relocating logical information or restarting computation.

\emph{Algorithmic Overview:} The framework employs a defect-detection strategy using DBSCAN clustering over syndrome mismatches to identify faulty regions during execution. Upon locating a defect, the surrounding stabilizers are merged into a superstabilizer encircling the faulty patch. Two variants are used: a basic approach where the superstabilizer is directly inferred, and a shell-based approach that uses repeated gauge operator measurements for higher fidelity. The decoding graph is updated accordingly to incorporate the new stabilizer geometry.

While this method does not recover the original code distance, it avoids full logical block discard and circumvents the need for preallocated redundant logical qubits. For example, when applied to a 3×3 defect in a surface code of distance $d=15$, the adaptive patch deformation preserved logical functionality with a threshold only marginally lower than the defect-free baseline (2.7\% vs. 2.9\%), using fewer qubits than static overprovisioning would require. Thus, it reduces qubit overhead by avoiding global redundancy in favor of localized code deformation.

\end{svgraybox}

\begin{svgraybox}
\textbf{Case Study 3.1.2: Surf-Deformer \cite{yin2024surfdeformermitigatingdynamicdefects}}

Surf-Deformer introduces a composable instruction set that enables fine-grained patch resizing and defect excision through gauge fixing. Unlike the adaptive surface code framework, it not only removes defective qubits but also restores lost code distance and supports dynamic routing.

\emph{Algorithmic Overview:} The core technique models patch manipulation as a set of gauge-preserving transformations: stabilizer-to-gauge (S2G), gauge-to-stabilizer (G2S), stabilizer-to-stabilizer (S2S), and gauge-to-gauge (G2G) operations. At runtime, the controller applies $PatchQ\_RM$ to remove defective zones and $PatchQ\_ADD$ to regrow the patch boundary. A layout generator estimates inter-patch spacing ($\Delta d$) based on defect rate and expected growth, allowing compilation-time resource planning. Read Section IV from \cite{yin2024surfdeformermitigatingdynamicdefects} for a detailed understanding.

Compared to Q3DE \cite{Suzuki_2022}, which doubles the patch size to recover distance, Surf-Deformer achieves equivalent error suppression with 2× fewer qubits. On QFT-100-20 benchmarks under burst faults, Surf-Deformer reduced retry risk by up to 6× compared to the adaptive surface code framework while consuming ~50\% fewer physical qubits than Q3DE. Its ability to dynamically restore code integrity while preserving communication links makes it highly efficient for fault-aware layouts.

\end{svgraybox}

\begin{svgraybox}
\textbf{Case Study 3.1.3: Quantum Prometheus \cite{qprom}}

This framework proposes a compile-time strategy for reducing ancilla overhead by reusing the same ancilla qubits to measure both $X$- and $Z$-type stabilizers, scheduled in interleaved sub-rounds instead of simultaneously.

\emph{Algorithmic Overview:} The method divides each QEC round into two sequential sub-rounds. In the first, $X$-stabilizers are measured using a fixed set of ancilla qubits; after reset, the same ancillas are reused to measure $Z$-stabilizers in the second sub-round. This reuse is compiler-inserted and does not require geometric patch modification. Logical decoding is applied only after a full round, preserving syndrome integrity.

This simple reuse strategy surprisingly reduces ancilla requirements by 50\% and achieves a total qubit reduction of ~25\%. Crucially, for large code distances ($d \geq 13$), a modified surface code of distance $d+2$ using this framework consumes fewer qubits than a standard code of distance $d$, while achieving better logical error suppression. For instance, a $d=31$ modified surface code achieved a $1.5 \times 10^{-27}$ logical error rate with 1441 qubits, compared to a $d=27$ original code requiring 1457 qubits for a higher error rate of $4.27 \times 10^{-24}$.

\end{svgraybox}

\subsection{$T$-gate Optimization}

In two‐dimensional topological codes such as the surface code, all logical Clifford operations ($H$, $S$, $CNOT$) admit low‐overhead, transversal implementations or lattice‐surgery realizations \cite{Litinski_2019}; by contrast, the non‐Clifford $T$‐gate (a $\pi/8$ phase rotation) is \emph{not} transversal \cite{ghosh2025survivaloptimizedevolutionaryapproach} and must be effected by injecting a high‐fidelity “magic” state into the code via a dedicated lattice‐surgery protocol. Each such injection consumes one distilled
\(
    \lvert m \rangle \;=\; \frac{\lvert 0\rangle + e^{i\pi/4}\lvert 1\rangle}{\sqrt{2}},
\)
magic state , and because raw injections at physical error rate $p_{\mathrm{phys}}$ remain faulty at $\mathcal{O}(p_{\mathrm{phys}})$, recursive magic‐state distillation is required to reach logical error rates below $10^{-10}$ for classically intractable algorithms \cite{Litinski_2019_msd}.
Two natural resource metrics then arise:
\begin{itemize}[label={--}]
  \item \textbf{$T$-Count}: the total number of T (and T$^\dagger$) gates in the circuit, each corresponding to one $\lvert m\rangle$ consumption and thus setting the \emph{volume} (qubit footprint) of distillation factories \cite{ghosh2025survivaloptimizedevolutionaryapproach}\cite{Amy_2014}\cite{fast_todd}\cite{todd} \cite{pyzx}.
  \item \textbf{$T$‐Depth}: the minimum number of sequential layers of commuting T‐gates; because each layer incurs a full round of magic‐state injection and surface‐code syndrome extraction (on the order of one code cycle), T‐depth directly controls the \emph{latency} of the logical operation\cite{ghosh2025survivaloptimizedevolutionaryapproach}\cite{Amy_2014}\cite{fast_todd}\cite{todd} \cite{pyzx}.
\end{itemize}
Moreover, each distilled $\lvert m\rangle$ must itself be error‐corrected by a distance–$d$ patch, occupying $d\times d$ data qubits plus ancillas, so that achieving high fidelity often requires hundreds of magic states in parallel—amounting to tens or hundreds of thousands of physical qubits devoted solely to distillation.
Thus, any reduction in \emph{T‐count} directly shrinks the factory footprint, while any collapse of \emph{T‐depth} proportionally reduces the number of distillation rounds (and hence overhead qubits) needed. This key optimization can be performed both in the early-stage circuit synthesis by exploiting polynomial methods and rewriting gates graphically, as well as in later, post-Clifford pruned stages of circuit synthesis by merging $T$-gate layers that commute mathematically.

\subsubsection*{Early‐Stage Circuit Synthesis}

Since each logical $T$‐gate requires its own distilled magic state, and hence a full round of surface code syndrome extraction, any reduction in the $T$‐count or $T$‐depth before backend packing directly shrinks both the qubit footprint of magic‐state factories and the overall execution latency.  Early‐stage (front‐end) synthesis, therefore, seeks to eliminate non‐Clifford rotations as early as possible, at the algebraic, graphical, or tensor‐network level, so that downstream scheduling and resource allocation never “see” redundant $T$‐gates. We present a taxonomy of the current state-of-the-art circuit synthesizers specifically designed to reduce the $T$-gates.

\begin{itemize}[label={--}]
  \item \emph{Phase‐Polynomial Solvers: }  
    Extract the CNOT+$T$ subcircuit into a Boolean phase polynomial 
    \(\phi:\mathbb Z_2^n\to\mathbb Z_8\) plus a gate‐synthesis matrix,  
    then formulate T‐depth minimization as a matroid‐partition problem.  
    Representative tools include Tpar \cite{Amy_2014}, TODD \cite{todd}, and Faster‐TODD\cite{fast_todd}.
  \item \emph{ZX‐Calculus Rewriters: }  
    Translate the circuit into a graph‐like ZX‐diagram \cite{zx} of Z‐spiders and Hadamard edges,  
    apply local complementation and gadget‐fusion rules to collapse phase gadgets,  
    and perform phase‐teleportation to eliminate redundant \(\pi/4\) rotations.  
    Representative tools include PyZX used for $T$-gate optimization \cite{pyzx}.
  \item \emph{Tensor‐Network Unitary Synthesis.}  
    Represent single‐qubit gate sequences as a compact tensor network (MPS),  
    compute overlaps with the target unitary via efficient tensor contractions,  
    and sample high‐quality sequences under a T‐budget.  
    A representative tool is TRASYN \cite{hao2025reducingtgatesunitary}.
\end{itemize}

All three frameworks isolate a residual “phase gadget” structure, whether as third‐order tensor entries 
\(\mathcal T_{ijk}=\frac{\partial^3\phi}{\partial x_i\,\partial x_j\,\partial x_k}\),  
as degree‐one spiders in a ZX‐diagram, or as MPS tensors indexing gate blocks.  
Optimization then reduces to a combinatorial problem—low‐rank tensor decomposition or minimum matroid cover for T‐depth, graph rewrites for T‐count, or sampling from a tensor network under T‐budget. Therefore, on integrating the circuit synthesis ideas in the QEC design workflow, it looks somewhat like this:

\begin{enumerate}
  \item \emph{Decomposition:} parse the user’s target unitary into phase‐polynomial, ZX, and tensor‐network forms.
  \item \emph{Optimization Passes:} run the matroid‐partition solver, apply ZX rewrite rules, and/or perform tensor‐network sampling.
  \item \emph{Extraction:} reconstruct minimized Clifford+$T$ (or Clifford+\(U_3\)) circuits from each optimized representation.
  \item \emph{Selection and Hybridization:} compare resulting $T$‐count, $T$‐depth, and ancilla requirements; optionally chain or interleave passes (e.g.\ PyZX followed by TODD and/or methods in \cite{pyzx}).
\end{enumerate}

This automated early-stage pipeline ensures that the bulk of expensive magic‐state resources are never allocated to redundant rotations and provides a rich design space of algebraic, graphical, and tensor‐network “knobs” to tune for any hardware and fidelity constraint.

\begin{svgraybox}
\textbf{Case Study 3.2.1: Tpar \cite{Amy_2014}}

The $T$-depth optimization process for $CNOT$+$T$ circuits proceeds through four key steps. First, in the \emph{phase-polynomial extraction} phase, the $CNOT$+$T$ subcircuit is identified and its corresponding Boolean phase polynomial is computed as 
\[
\phi(\mathbf{x}) = \sum_{i=1}^L c_i\,f_i(\mathbf{x}), \quad c_i\in\mathbb{Z}_8,\; f_i:\{0,1\}^n\to\{0,1\}.
\]
Next, during \emph{matroid construction}, a matroid $(S,\mathcal{I})$ is formed over the set $S=\{(c_i,f_i)\}$, with the collection of independent sets defined by
\[
\mathcal{I} = \left\{T\subseteq S \;\middle|\; \dim(V) - \mathrm{rank}(T) \le n - |T| \right\}, \quad V = \mathbb{F}_2^n.
\]
The third stage involves \emph{partitioning} this set using Edmonds’ algorithm, which splits $S$ into the minimum number $k$ of independent subsets:
\[
S = I_1 \cup I_2 \cup \cdots \cup I_k,\quad I_j\in\mathcal{I}.
\]
Finally, the \emph{circuit re-synthesis} step is performed for each subset $I_j$: a $CNOT$ network is synthesized via Gaussian elimination to compute all $f_i \in I_j$, followed by a single layer of $T$-gates corresponding to the coefficients $c_i$, and then the $CNOT$ network is uncomputed. This results in a quantum circuit with $T$-depth exactly $k$, which is provably minimal under the given matroid-based model.

On $m$-bit Galois-field multipliers, the $T$-depth was reduced by approximately $79\%$ and the $T$-count by approximately $42\%$ without the use of ancillas. When $O(L)$ ancillas were allowed, the $T$-depth collapsed from $12(m - 1)$ layers to just $2$ layers. For $\Lambda_k(X)$ gates, the $T$-count was reduced from $7(4k - 8)$ to $3(4k - 8) + 4$, and the $T$-depth improved from $3(4k - 8)$ to $4k - 8$ using $k - 2$ ancillas.

\end{svgraybox}

\begin{svgraybox}
\textbf{Case Study 3.2.2: ZX‐Calculus Rewriting with PyZX \cite{pyzx}\cite{todd}\cite{fast_todd}}

The optimization process based on ZX-calculus proceeds through a sequence of structured transformations. First, in the \emph{ZX-diagram translation} step, the Clifford+$T$ circuit is converted into a graph-like ZX-diagram composed of Z-spiders connected via Hadamard edges. Next, during \emph{Clifford simplification}, local complementation and Pauli-spider removal are applied to eliminate all residual Clifford structure. In the \emph{phase-gadget fusion} stage, Identity Fusion and Gadget Fusion rules are applied iteratively to merge $\pi/4$ phase gadgets that share identical neighborhoods. This is followed by \emph{phase teleportation}, where symbolic phase variables are tracked during fusion; each fusion zeroes one variable and sums its phase into another, thereby recording cancellations. After this, the \emph{residual matroid pass} \cite{todd} is performed by extracting the updated phase polynomial and invoking the TODD algorithm to partition the remaining terms into layers of minimal $T$-depth. Finally, a \emph{faster enhancement} \cite{fast_todd} step improves the efficiency of the residual solver through advanced rank-decomposition routines, enabling the handling of larger sets of phase terms with improved computational performance.

PyZX alone matches or improves the then state-of-the-art (Tpar and/or TODD) $T$-count on approximately $72\%$ of reversible-arithmetic benchmarks, achieving up to a $50\%$ reduction on circuits rich in Hadamard gates. Incorporating the TODD algorithm provides an additional average $T$-count reduction of approximately $15\%$, thereby enabling efficient synthesis of circuits with more than $10^4$ phase terms.

\end{svgraybox}

\begin{svgraybox}
\textbf{Case Study 3.2.3: TRASYN \cite{hao2025reducingtgatesunitary}}

TRASYN focuses on the synthesis of the Clifford +$U3$ gate set, aiming to reduce the overall $T$-count as well as the Clifford gate count in the resulting circuit. The gate synthesis method proceeds as follows. First, a \emph{gate-block library} is precomputed, consisting of tensor-network representations of short sequences of Clifford+\(U_3\) operations. Next, the target single-qubit unitary—or a small multi-qubit block—is encoded as a Matrix Product State (MPS) by evaluating its overlap with elements of the gate-block library. Then, under a \emph{T-budget} constraint, tensor-network contractions are used to compute fidelity overlaps for candidate sequences, and sampling is performed to identify those sequences that satisfy the prescribed T-count limit. Finally, the \emph{sequence extraction} step selects the sequence with the highest fidelity and translates it into a native Clifford+\(U_3\) circuit. This process yields implementations that require fewer T gates than standard \(R_z\)-only decomposition techniques.

For arbitrary single-qubit rotations, TRASYN achieves up to $2\times$ fewer $T$-gates compared to conventional three-gate \(R_z\) decompositions, while maintaining equivalent error bounds. On benchmark sets consisting of random unitaries, the average $T$-count is reduced by $30\text{–}50\%$, with end-to-end synthesis times of under one second per rotation. 

\end{svgraybox}

\subsubsection*{Late-Stage Circuit Synthesis}

At this stage of the design flow, all Clifford gates have been commuted to the end of the circuit or absorbed into measurement bases \cite{Litinski_2019}\cite{Litinski_2019_msd}, leaving a sequence of non‐Clifford $\pi/8$ rotations organized into $m$ layers:
\[
  C \;=\; L_1 \,\to\, L_2 \,\to\, \cdots \,\to\, L_m,
  \quad
  L_j = \exp\Bigl(i\frac{\pi}{8}\sum_{q=1}^n P_{q}^{(j)}\Bigr),
\]
where each $P_{q}^{(j)}\in\{I,X,Y,Z\}$ is a Pauli operator on qubit $q$.  Because each layer $L_j$ requires a fresh batch of distilled magic states and a full round of surface‐code syndrome extraction, the total T‐depth $m$ dictates both the number of distillation rounds and the overall latency of the logical computation.
The objective of late‐stage synthesis is to minimize $m$ by merging layers whose Pauli‐product rotations commute, without introducing additional $T$-gates.  Since the search space of all possible mergings grows combinatorially with $m$, practical approaches employ heuristic or meta‐heuristic search to identify near‐optimal packings of commuting layers.

\begin{svgraybox}

\textbf{Case Study 3.2.4: Genetic‐Algorithm for $T$-depth optimization \cite{ghosh2025survivaloptimizedevolutionaryapproach}}

In the GA-based approach for optimization in the late-circuit synthesis, the post‐Clifford circuit is represented as a list of length‐$n$ Pauli vectors $\{L_j\}_{j=1}^m$.  Optionally, high‐density layers may be split into finer sub‐columns along low‐density qubit subsets to expose additional merge candidates. Compute for each layer $L_j$ its T‐count $T_j$ and density $D_j = T_j / n$.  For every pair $(i,j)$ whose Pauli products commute, assign a score
\[
  \mathrm{score}(i,j) \;=\; 1 - \lvert D_i - D_j\rvert 
    \;+\; \beta\,\bigl(T_{\max} - (T_i + T_j)\bigr),
\]
with tunable weight $\beta\in[0,1)$.  Select a maximal non‐overlapping matching of highest‐scoring pairs to seed the Genetic‐Algorithm population.
\textbf{Genetic‐Algorithm Framework: }
The layer-collapsing algorithm proceeds through an evolutionary optimization framework. It begins with an \emph{initialization} step, wherein an initial population of merge-sets is created; each merge-set is a disjoint collection of commutable gate pairs. Next, during \emph{fitness evaluation}, the fitness of each individual is defined as the total number of merges it contains. The \emph{selection and elitism} phase retains the top $k$ individuals without modification and generates new offspring through crossover, by combining merge-sets, and occasional mutation, which resets portions of the chromosome. Finally, in the \emph{iteration} phase, the best merge-set is applied to collapse the paired circuit layers into single layers. The circuit representation is then updated, and the process is repeated until no further merges are possible.

On synthetic circuits with 90–100 qubits and maximal T-density, the genetic algorithm reduced $T$-depth by up to $79.2\%$, e.g., from approximately $\sim\!1024$ layers to $\sim\!214$—and $T$-count by $41.9\%$, outperforming lookahead heuristics by a factor of approximately $2.6\times$ in $T$-depth reduction. On standard reversible-arithmetic benchmarks \cite{Amy_2014}\cite{pyzx}, such as adders and Galois-field multipliers, the average T-depth decreased by a factor of $2.58\times$ when compared to zero-ancilla matroid-based methods, all achieved without the use of additional ancilla qubits.

\end{svgraybox}

\subsection{Optimizing the Surface Code Layout}

\subsubsection*{Distillation Protocols in Surface-Code Architectures}
Magic-state distillation in the surface code can be mapped onto \emph{distillation blocks} \cite{Litinski_2019}\cite{Litinski_2019_msd}, each of which implements a fault-tolerant multi-qubit Pauli product measurement via lattice surgery on patches of code distance \(d\).  A distillation block consists of:
\begin{itemize}[label={--}]
  \item Logical qubit patches of size \(d\times d\),  
  \item Ancillary patches for multi-patch Pauli product measurements (to realize \(\pi/8\) rotations),  
  \item Classical feed-forward paths for measurement outcome processing.  
\end{itemize}
The geometric cost of each block is measured by its space–time volume
\[
V \;=\; A \times T,
\]
where \(A\) is the number of \(d\times d\) tiles and \(T\) the number of code cycles.  For example, a single-level 15-to-1 protocol occupies an \(11d\times5d\) tile region over \(12d\) cycles, yielding
\[
V_{15\to1}\;=\;660\,d^3.
\]
Alternative protocols include:
\begin{itemize}[label={--}]
  \item \textbf{20-to-4:} occupies \(\approx14d^2\) tiles over \(4d\) cycles, producing four outputs with \(\mathcal{O}(p^2)\) error \cite{Litinski_2019_msd},
  \item \textbf{Higher-rate codes} (e.g.\ 116-to-12, 225-to-1): trade larger \(A\times T\) for multi-output or improved error scaling.
\end{itemize}
In a full-scale computation, the data-block’s demand—set by the circuit’s \(T\)-count and \(T\)-depth—must be matched by the throughput of parallel distillation blocks. The key trade-offs in magic state distillation and factory design are as follows. \emph{Underprovisioning} leads to execution stalls due to insufficient availability of distilled states. In contrast, \emph{overprovisioning} results in increased qubit overhead, consuming more hardware resources than necessary. Additionally, while \emph{higher-level} distillation protocols can reduce the per-state cost in terms of $T$-gates or time, they typically require a larger factory.

\begin{svgraybox}

\textbf{Case Study 3.3.1: Protocol Selection by Workload \cite{Litinski_2019_msd}}

We analyze three representative \(T\)-gate workloads (physical error rate \(p=10^{-4}\)) and identify the protocol minimizing space–time cost per magic state.

\begin{center}
\begin{tabular}{l|cc|lc}
\bfseries Workload & \bfseries \(T\)-count & \bfseries \(T\)-depth & \bfseries Best Protocol & \bfseries Cost per state \\\hline
High-rate streaming    & \(10^8\)  & \(10^6\)  & 20-to-4                & \(27\,d^3\) \\
Ultra-low error        & \(10^{10}\)& \(10^4\) & 15-to-1 \(\times\) 15-to-1 & \(25.9\,d^3\) \\
Resource-constrained   & \(10^6\)  & \(10^6\)  & 15-to-1                & \(6.3\,d^3\) \\
\end{tabular}
\end{center}

\noindent\textbf{High-rate streaming.}  
When \(T\)-count \(\gg\) \(T\)-depth, multi-output blocks (e.g.\ 20-to-4) maximize throughput per tile—ideal for pipelined factories.

\noindent\textbf{Ultra-low error.}  
Moderate depth with stringent fidelity demands favors cascading two 15-to-1 stages, achieving \(\mathcal{O}(p^9)\) error at \(\approx25.9\,d^3\) per output.

\noindent\textbf{Resource-constrained.}  
Low-depth, small-scale machines use a single-level 15-to-1 block to minimize footprint (\(\approx6.3\,d^3\)) at the cost of higher latency.
These examples demonstrate that matching protocol choice to \((T\text{-count},T\text{-depth})\) yields orders-of-magnitude reductions in space–time overhead.
\end{svgraybox}

\begin{svgraybox}
\textbf{Case Study 3.3.2: Analysis of algorithms to choose Magic State Distillation Protocols \cite{chatterjee2025qspellbookcraftingsurfacecode}}

Each magic state distillation (MSD) protocol is characterized by three primary metrics: the number of surface-code tiles \(D_p\) required, the number of distillation steps \(S_p\), and the number of output magic states \(k_p\) produced per round.  For a given physical error rate \(p\), the success probability of protocol \(p\) is \(P_s = (1 - p)^{N_p}\), where \(N_p\) is the total number of raw noisy states consumed.  Consequently, the effective latency per distilled state is expressed as \(\frac{S_p}{k_p\,P_s}\).  Representative examples include the 15-to-1 protocol with \(D_p=11\), \(S_p=11\), \(k_p=1\), and the 20-to-4 protocol with \(D_p=14\), \(S_p=17\), \(k_p=4\).
The Brute Force algorithm exhaustively enumerates all sequences of distillation rounds up to a fixed length \(L\), simulating resource consumption and pipeline stalls to compute exact total tile-time products.  Despite its optimality guarantee, the exponential search complexity (\(O(|P|^L)\)) renders it impractical for large \(L\).  Dynamic Programming (DP) mitigates this by defining a cost function \(C(i,s)\) over protocol index \(i\) and cumulative step \(s\), and applying the recurrence
\[
C(i,s) = \min_{p\in P}\bigl\{C(i-1,\,s-S_p) + D_p\bigr\}\,,
\]
yielding an exact tile-minimal solution in time \(O(L\cdot S_{\max}\cdot |P|)\).  **Greedy** selection iteratively chooses the protocol \(p^* = \arg\min_p\!\bigl(\tfrac{D_p}{k_p} + S_p\bigr)\), updating available magic-state counts and accounting for stalls; this yields near-optimal latency with linear complexity \(O(L\cdot|P|)\).  A Random baseline assigns a fixed protocol per optimization objective without adaptation.

When minimizing tile usage, DP reproduces the Brute Force optimum with 0\% tile overhead, whereas Greedy incurs an average overhead of 41.3\% and Random 14.7\%.  For latency minimization, Greedy deviates by only 7.6\% from the Brute Force optimum, substantially outperforming DP (64.6\% overhead) and Random (148.5\%).  In a balanced multi-objective scenario, DP and Greedy yield distinct Pareto-optimal configurations: DP points minimize tiles at the cost of extra steps, while Greedy points minimize steps with higher tile counts.

\end{svgraybox}

\subsection{ML in Decoder Automation}
Machine‐learning (ML) modules can be introduced into the QEC decoding pipeline at carefully chosen junctures to accelerate decision making and to capture complex noise correlations that classical decoders struggle to model.  At its core, decoding begins with the syndrome extraction process, which yields a binary vector
\[
\mathbf{s}\;\in\;\{0,1\}^m
\]
of length \(m\) per QEC cycle.  Traditionally, this vector is fed directly into a matching or propagation algorithm (e.g.\ MWPM, union‐find, belief propagation) to propose a recovery operator \cite{wang2023transformerqecquantumerrorcorrection}\cite{10682779}\cite{wang2024artificialintelligencequantumerror} \cite{Overwater_2022}\cite{gong2023graphneuralnetworksenhanced}.  In an automated flow, the first opportunity for ML intervention lies in \emph{syndrome preprocessing}, where one learns a mapping
\[
\mathbf{s}\;\longmapsto\;\mathbf{z}\;\in\;\mathbb{R}^d,\quad d\ll m,
\]
via an autoencoder or embedding network.  By compressing syndrome information into a low‐dimensional latent space \(\mathbf{z}\), one both reduces the memory footprint and highlights the most salient error features—whether spatially clustered bit‐flips or time‐correlated measurement faults—before any costly graph search.
Once a compact representation is available, the next “injection point” is the \emph{syndrome‐to‐recovery mapping} stage.  Here, one replaces, or augments, the classical lookup table (LUT) or brute‐force search by training a neural classifier \(f\) to predict the appropriate Pauli correction
\[
R_{\rm pred} = f(\mathbf{z})
\]
from among the set \(\mathcal{R}=\{I,X,Z,Y\}^{\otimes k}\) of \(k\) logical‐qubit operators.  This reframing transforms the combinatorial decoding problem into a constant‐time inference task, yielding an \(O(1)\) decision latency per cycle.  In practice, a shallow feed‐forward or convolutional network can achieve logical‐error rates on par with minimum‐weight perfect matching for small code distances, but with vastly reduced—and hardware‐friendly—runtime.
Beyond single‐round decoding, real QEC experiments demand resilience to \emph{temporal correlations} arising from repeated measurements.  By treating the syndrome record as a time series \(\{\mathbf{s}_t\}_{t=1}^T\), one can employ sequence models, e.g., attention‐based encoders, to distinguish between transient measurement flips and persistent data errors \cite{wang2023transformerqecquantumerrorcorrection}. The result is a decoder that adaptively weights historical syndrome events when proposing corrections, yielding enhanced suppression of correlated error strings without sacrificing the modular structure of the classical decoder.
Finally, practical deployment on FPGA or ASIC platforms imposes stringent area and latency constraints. \emph{Resource‐aware model compression} (via pruning or quantization) and \emph{co‐design} of neural architectures with digital logic pipelines ensure that syndrome acquisition, ML inference, and Pauli‐frame updates form a seamless, low‐latency hardware flow.  Hybrid schemes—such as learned edge‐weight prediction within a classical matching algorithm, preserve polynomial‐time guarantees while infusing context‐sensitive intelligence into each stage of the decoder.

\begin{svgraybox}
\noindent\textbf{Case Study 3.4.1: Multi-layer Perceptron (MLP) in QEC Decoding \cite{Overwater_2022}}

A pure-error peel-off routine deterministically removes stabilizer contributions from the raw syndrome, after which a shallow MLP infers the remaining logical Pauli correction.

\textbf{MLP Design \& Training: }
The neural classifier takes as input the $d^2-1$ independent defect bits remaining after pure-error cancellation for a distance-$d$ surface code, and outputs two logits $\ell_X,\ell_Z$ corresponding to an uncompensated logical $X$ and $Z$ flip.  The network comprises two fully connected hidden layers of size 256 and 64, each followed by a square-nonlinear (SQNL) activation.  Training data are sampled on-the-fly under a depolarizing noise model at physical error rate $p\approx p_{\rm pseudo}$ (the MWPM pseudo-threshold for each $d$), ensuring coverage of hard, high-weight error patterns.  Parameters $\theta$ are optimized via ADAM to minimize

\begin{align*}
    \mathcal{L}(\theta) \;=\; -\mathbb{E}_{(\mathbf{s},y)}\Bigl[y_X\log\sigma(\ell_X(\mathbf{s})) + (1-y_X)\log\bigl(1-\sigma(\ell_X(\mathbf{s}))\bigr) + \\ 
    y_Z\log\sigma(\ell_Z(\mathbf{s})) + (1-y_Z)\log\bigl(1-\sigma(\ell_Z(\mathbf{s}))\bigr)\Bigr] + \\ \lambda\|\theta\|_2^2 + \gamma\,Q(\theta)
\end{align*}
where $y_X,y_Z\in\{0,1\}$ are the true logical flips, $\sigma$ is the sigmoid, and $Q(\theta)$ is a quantization-penalty driving weights toward small integer levels.

The logical-error rate $p_L(d,p)$ is defined as the probability that the combined pure-error + MLP correction yields $CE\notin\mathcal{S}$.  In the low-$p$ regime, one observes the characteristic scaling
\[
p_L(d,p)\;\approx\;A\,p^{\frac{d+1}{2}}\,,\quad d=3,5,7,9,
\]
with an exponent matching that of MWPM. The MLP high-level decoder achieves a pseudo-threshold $p_{\rm th}\approx1.2\%$ (comparable to blossom matching) and maintains the $O(1)$ inference cost per cycle.  Incorporating fourfold rotational weight-sharing reduces the variance in $p_L$ across random noise realizations and slightly increases the exponent coefficient $A$ at larger $d$.  Under uniform $b$-bit quantization of weights ($b=3\text{–}7$), the threshold degrades by only $\sim0.1\%$, demonstrating that
\[
p_L(d,p; b\hbox{-bit}) - p_L(d,p; \infty\hbox{-bit}) \;=\; O\bigl(p^{\tfrac{d+1}{2}}\bigr)
\]
remains small even for aggressive finite-precision constraints.

\textbf{Hardware Implementation: }
A custom ASIC implementation of the MLP inference stage achieves sub-10 ns latency per cycle—well within a typical 440 ns error-correction window—and occupies less than 0.1 mm$^2$ at $d=7$.  On FPGA prototypes, pipelined execution of the two-layer network and the pure-error XOR tree sustains a $>100$ MHz decoding clock.  Power consumption scales as $\Theta(d^2)$, indicating that beyond $d\approx11$, resource trade-offs must be carefully balanced against fidelity gains.
\end{svgraybox}

\begin{svgraybox}
\noindent\textbf{Case Study 3.4.2: Graph Neural Networks (GNNs) in QEC Decoding \cite{gong2023graphneuralnetworksenhanced}}\\
Leveraging the detector graph topology via a GNN enables direct learning of spatial–temporal error correlations beyond local matching heuristics.

\textbf{GNN Design \& Data Representation: }
A space–time graph $G=(V,E)$ is constructed from syndrome events: each node $i$ holds features $(b_i,x_i,y_i,t_i)$ encoding stabilizer type and coordinates; edges connect nearest neighbors within radius $r$, weighted by $e_{ij}=1/\sqrt{\Delta x^2+\Delta y^2+\Delta t^2}$ and pruned to max degree $k$. Seven graph-convolutional layers
\[
\mathbf{h}_i^{(l+1)} = \mathrm{ReLU}\bigl(W_1\mathbf{h}_i^{(l)} + W_2\sum_{j\in\mathcal{N}(i)} e_{ij}\,\mathbf{h}_j^{(l)} + b\bigr)
\]
A mean-pooling $\mathbf{h}_G = |V|^{-1}\sum_i\mathbf{h}_i^{(L)}$ feeds two MLP heads with sigmoid outputs for logical-flip probabilities $\hat p_X,\hat p_Z$. 
On training on Stim \cite{gidney2021stim} and experimental data for repetition codes at $d=3,5,25$ using Adam on binary cross-entropy.  On surface-code simulation ($p=10^{-3}$, $d\le9$), the GNN achieves logical-error rate scaling
\[
p_L(d,p)\approx A\,p^{(d+1)/2}
\]
with threshold $p_{\rm th}\approx1\%$, outperforming MWPM in low-$d$ regimes and matching its exponent.  Inference complexity grows as $O(|V|+|E|)\approx O(d^2t)$, substantially below the $O(d^3)$ of blossom matching. 

\textbf{Hardware Implementation: }
While GPU-accelerated implementations sustain real-time decoding for $d\le9$, ASIC or FPGA deployment requires fixed-size, quantized GNN kernels and sparse message-passing units to meet sub-100 ns latency per QEC cycle.  Edge-pruning and degree capping are critical to bounding on-chip memory and compute.  
\end{svgraybox}

\begin{svgraybox}
\noindent\textbf{Case Study 3.4.3: Transformers in QEC Decoding \cite{wang2023transformerqecquantumerrorcorrection}}\\
By leveraging the global receptive field of self-attention, the approach jointly processes space–time syndrome information, facilitating effective transfer learning across different quantum code distances.
Syndromes are encoded on a $(D+1)^2\times(D+1)\times T$ cubic grid with 6-dimensional feature vectors ($X/Z$ check flags, syndrome bits, temporal markers). Tokens are projected to embeddings, 3-dimensional sinusoidal positional encodings are added, and then flattened to a 1-dimensional sequence.  A Transformer encoder of $L$ layers (MHSA + FFN) yields syndrome representations.  A Transformer decoder ingests data‑qubit positional embeddings and attends to encoder outputs, producing per‑qubit error logits.  Finally, an auxiliary global parity head pools encoder features to predict overall syndrome parity. 
It is trained on simulated and experimental datasets at distance $d=5$ using a mixed loss
\[
\mathcal{L} = \mathcal{L}_{\rm local} + \alpha\,\mathcal{L}_{\rm global},
\]
where $\mathcal{L}_{\rm local}$ is binary cross‐entropy per data qubit and $\mathcal{L}_{\rm global}$ from pooled parity. Adam is used with warmup + cosine decay for 100 epochs.  For new distances ($d=7,9,11,13$), fine‑tuned pretrained weights for 10 epochs at reduced learning rate, saves $>10\times$ training cost. The logical‑error rate $p_L(d,p)$ is evaluated under a phenomenological noise model for distances $3\le d\le13$ and $p=10^{-3}\text{–}5\times10^{-2}$.  Transformer‑QEC achieves thresholds $p_{\rm th}\approx3.8\%$ and consistently outperforms UF, MWPM \cite{wu2023fusionblossomfastmwpm}, and MLP\cite{Overwater_2022} across all benchmarks.  The logical scaling follows
\[
p_L(d,p)\sim A\,p^{\lfloor(d+1)/2\rfloor},
\]

matching theoretical exponents while maintaining $O(N)$ attention complexity (with $N\approx D^2T$). Transfer learning further reduces the on-chip training overhead by an order of magnitude.

\end{svgraybox}

\subsection{Formal Verification in QEC}

The development of large-scale FTQC systems depends critically on the correctness of the QEC infrastructure \cite{Huang_2025}. While simulation-based testing remains useful for empirical exploration, it fails to provide formal guarantees of correctness or to scale to systems involving hundreds or thousands of physical qubits. Formal verification offers a principled alternative: a mathematically rigorous methodology to prove that QEC codes and operations preserve the intended logical behavior despite the presence of noise and implementation imperfections.
A robust verification methodology for QEC codes is generally structured around three foundational pillars: a domain-specific language for encoding QEC procedures, a formal semantics and logical assertion framework, and a deductive proof system for establishing correctness properties \cite{wu2021qecvquantumerrorcorrection}.
\begin{itemize}[label={--}]
    \item \textbf{Domain-Specific Representation Language: }  
    The first requirement is a structured language capable of expressing the components of a QEC protocol, viz., encoding, stabilizer measurements, syndrome-dependent branches, recovery procedures, and logical gate implementations. The language must support quantum data operations alongside quantum-classical control constructs such as conditionals and loops. Also, the language should allow stabilizers to be treated as primary objects, abstracting the hardware-level details of measurement circuits while enabling logical reasoning over code behavior. This abstraction makes it possible to model code families like repetition and surface codes independently of specific architectures.
    \item \textbf{Formal Semantics and Logical Assertions: }  
    To reason about correctness, the language must be paired with a well-defined semantic model. Operational semantics describe how program states evolve during execution, typically over a combination of quantum states and stabilizer configurations. Denotational semantics provide a higher-level view of program behavior, enabling reasoning about loops and composition via fixed-point interpretations. Assertions are specified using logical predicates defined over stabilizers or Pauli expressions, serving as preconditions and postconditions for program operations. This approach avoids the intractable cost of general Hermitian-based predicate reasoning while preserving the ability to express meaningful correctness conditions for stabilizer codes.
    \item \textbf{Deductive Proof System for QEC Programs: }  
    The final component is a sound, syntax-directed proof system for deriving correctness judgments over QEC programs. Structured as a quantum analog of Hoare logic \cite{hoare}\cite{fifty_hoare}, the system defines how assertions are propagated through each language construct—initializations, unitary operations, stabilizer measurements, branching, and loops. The logic supports compositional reasoning, enabling the decomposition of large programs into verifiable subcomponents. Importantly, for operations within the Clifford group, the stabilizer-based formulation yields polynomial-time verification procedures, in contrast to the exponential complexity required by general-purpose quantum verification tools.

\end{itemize}

These aspects of formal verification with respect to QEC allow rigorous certification of code behavior under fault models, support reasoning about logical gate implementations and syndrome-based corrections, and enable scalable verification workflows for realistic quantum codes. 

\begin{svgraybox}
\textbf{Case Study 3.5.1: QECV \cite{wu2021qecvquantumerrorcorrection}}

\textbf{Language Design: }  
The QECV framework introduces a domain-specific programming language, referred to as \texttt{QECV-Lang}, explicitly designed to model QEC routines using the stabilizer formalism as the core abstraction. In this language, stabilizers are treated as first-class program variables, and quantum operations are expressed in a high-level syntax that encompasses initialization, Clifford unitaries, Pauli measurements, and quantum-classical control constructs.
Each program is composed of statements such as:
\[
\texttt{q := }\ket{0}, \quad \texttt{s := }+Z_1Z_2, \quad \texttt{if } M[s] \texttt{ then } P_1 \texttt{ else } P_2,
\]
where \(s \in \mathcal{S}_n\) denotes a stabilizer expression over \(n\) qubits. This abstraction avoids low-level circuit specification and enables formal reasoning over logical operations, syndrome extraction, and feedback control. A companion assertion language, \texttt{QECV-Assn}, defines predicates over stabilizer relations, allowing properties of the encoded state to be captured compactly.

\textbf{Encoding Procedures: }  
Using \texttt{QECV-Lang}, canonical QEC protocols such as the 3-qubit repetition code and distance-3 surface code are encoded in a compositional manner. For example, the 3-qubit bit-flip code is described via the following high-level steps:
\begin{enumerate}
  \item Initialize physical qubits: \(\texttt{q1 := } \ket{0}, \texttt{q2 := } \ket{0}, \texttt{q3 := } \ket{0}\),
  \item Encode logical state with CNOTs: \(\texttt{q2 := CNOT(q1, q2), q3 := CNOT(q1, q3)}\),
  \item Define and measure stabilizers: \(s_1 = Z_1Z_2, s_2 = Z_2Z_3\),
  \item Apply syndrome-based correction: conditional application of \(X\) gates based on measured outcomes of \(s_1\) and \(s_2\).
\end{enumerate}
Stabilizer assignments and correction branches are represented symbolically and reasoned over collectively, enabling verification under all syndrome outcomes in a unified symbolic execution trace. The abstraction naturally extends to more complex codes, such as surface codes, where stabilizers are represented by local Pauli products over 2D lattices and syndrome extraction is modeled through iterative conditional logic.

\textbf{Proof Strategy: }  
The QECV framework defines an operational semantics over configurations \((\rho, \sigma)\), where \(\rho\) is a quantum state and \(\sigma\) maps stabilizer variables to Pauli strings. A corresponding denotational semantics defines the behavior of programs as superoperators acting on such configurations. The correctness of QEC programs is verified using a syntax-directed Hoare-style proof system over stabilizer assertions:
\[
\{A\} \; P \; \{B\},
\]
where \(A\) and \(B\) are assertions in \texttt{QECV-Assn}, such as logical stabilizers or error syndromes, and \(P\) is a \texttt{QECV-Lang} program.
Correctness proofs were constructed for logical initialization, Clifford gates, and error correction routines for both the repetition code and the surface code. For instance, in the 3-qubit code, the system verifies that an injected single-qubit \(X\) error is successfully corrected and the logical information is preserved:
\[
\{ X_i \cdot \mathcal{L} \} \; P_{\text{corr}} \; \{ \mathcal{L} \},
\]
where \(\mathcal{L}\) denotes the logical stabilizer and \(P_{\text{corr}}\) is the correction program.
Importantly, QECV demonstrates polynomial-time verification of these protocols, enabled by the restriction to stabilizer operations and Clifford semantics. This contrasts with prior frameworks requiring exponential resources to represent general quantum states and predicates. Empirical benchmarks confirm that verifying distance-\(d\) surface code routines scales as \(O(d^3)\), highlighting the tractability of the stabilizer-based formalism and its suitability for design automation in realistic fault-tolerant QEC pipelines.
\end{svgraybox}

\begin{svgraybox}
\textbf{Case Study 3.5.2: Verification of FTQC \cite{chen2025verifyingfaulttolerancequantumerror}}

\textbf{Language Design: }
Surface codes are a class of topological stabilizer codes defined on a planar lattice, where logical qubits are encoded in global degrees of freedom, and stabilizer checks are local and geometrically constrained. The error correction (EC) gadgets in surface codes consist of repeated syndrome measurements, ancilla-assisted Pauli measurements (e.g., using cat states or flag qubits), and classical feedback via decoders.
The behavior of an EC gadget is described as a classical-quantum program (\texttt{cq-prog}) \( S \), comprising quantum initializations, Clifford gates, mid-circuit measurements, classical branching, and repeat-until-success constructs. The fault-tolerance specification follows Definition 6(4): for any logical input state \( \ket{\psi} \), any Pauli error of weight \( r \), and any execution with \( s \) faults such that \( r + s \leq t \), the output must remain within the space \( \mathcal{S}_s(\ket{\psi}) \), where \( t = \lfloor \frac{d-1}{2} \rfloor \) is the code’s correctable threshold.

\textbf{Encoding Procedure: }
The symbolic execution framework encodes all cq-prog transitions as symbolic stabilizer updates. Each configuration is represented by a tuple:
\[
\langle S, \tilde{\sigma}, \tilde{\rho}, p, \varphi, \tilde{F} \rangle,
\]
where \( \tilde{\sigma} \) is the symbolic classical state, \( \tilde{\rho} \) is a symbolic stabilizer state parameterized over Boolean symbols, \( \varphi \) is the accumulated path condition, and \( \tilde{F} \) is a symbolic counter representing the number of injected faults.

Error injection is modeled symbolically by modifying stabilizer generators. For a symbolic state \( \tilde{\rho} = \langle (-1)^{g_1} P_1, \dots, (-1)^{g_m} P_m \rangle \), a fault injection on qubit \( q \) produces:
\[
\tilde{\rho}' = \langle (-1)^{g_1 \oplus e \cdot c_1} P_1, \dots, (-1)^{g_m \oplus e \cdot c_m} P_m \rangle,
\]
where \( e \) is a new fault symbol and \( c_i = 1 \) if \( P_i \) anti-commutes with the injected Pauli and 0 otherwise. This allows symbolic tracking of error propagation without enumeration.

At the end of each execution path, the symbolic output \( \tilde{\rho}' \) is compared with the ideal fault-free output \( \rho_{\text{ideal}} \). Fault-tolerance is certified if, for all satisfying valuations \( \mathbf{s} \), the Pauli distance \( D(\tilde{\rho}', \rho_{\text{ideal}}) \) is bounded:
\[
\forall \mathbf{s}, \quad \varphi(\mathbf{s}) \Rightarrow D(\tilde{\rho}', \rho_{\text{ideal}}) \leq \tilde{F}(\mathbf{s}).
\]
Here, \( D(\cdot, \cdot) \) computes the minimal weight of a Pauli operator transforming one stabilizer state into another. This constraint is encoded into SMT queries using symplectic vector representations of the Pauli group.

\textbf{Proof Strategy: }
The framework provides a complete and automated method for certifying fault-tolerance of surface code EC gadgets. For memory-less loops, the method is both sound and complete; for conservative loops, soundness is preserved under local non-propagation of errors. Verification produces either a formal certificate or a counterexample trace highlighting a violation of fault-tolerance.
This capability allows both design-time validation of error correction protocols and comparative analysis of EC strategies, such as differing ancilla schemes or decoder behaviors. The framework also supports symbolic abstraction of decoder logic, making it scalable to realistic surface-code gadgets with dozens of physical qubits.

\end{svgraybox}

%% file: author/section4.tex
\section{FTQC architectures}
\label{sec4}
The next step to establishing QEC codes is the integration of the entire design workflow in an FTQC framework. There have been recent studies showing how this can be achieved \cite{sparo} \cite{hetec}. Based on the existing research, we can say that any near-term FTQC architecture built upon surface codes should integrate the following structural layers:

\begin{itemize}[label={--}]
  \item \textbf{Logical‐Qubit Fabric: }
  A 2D lattice of distance-$d$ surface‐code patches, each comprising a $d\times d$ array of data qubits interleaved with ancilla qubits for $X$‐ and $Z$‐stabilizer extraction every QEC cycle. Logical Clifford operations are performed via lattice‐surgery merges and splits of adjacent patches, implementing $CNOT$, $H$, and $S$ gates with $O(d)$ time overhead and local syndrome updates, while the fabric’s native primitives for Pauli‐product measurements—realized through correlated ancilla circuits or twist‐defect sequences—enable arbitrary non‐Clifford rotations when combined with magic‐state injection.  

  \item \textbf{Magic‐State Factory: }
    Dedicated distillation blocks (e.g.\ Bravyi–Kitaev 15-to-1 magic state distillation protocol \cite{Litinski_2019_msd}) consume noisy $\lvert T\rangle$ injections and output states with error $O(p^{3})$, where $p$ is the physical gate error rate; these blocks occupy an $O(d)\times O(d)$ footprint of surface-code patches arranged in a pipelined sequence of distillation rounds to produce one high-fidelity $\lvert T\rangle$ every $O(d)$ code cycles; injection channels—surface-code ancilla chains—then ferry distilled $\lvert T\rangle$ states into the compute fabric with bounded routing latency.

  \item \textbf{Interconnects / Routing Layer: }
    A fault‐tolerant ancilla network, realized either as a fixed “bus” of qubits \cite{hetec} or as a mesh of surface‐code ancilla patches \cite{sparo}, supports logical‐state teleportation and multi‐qubit Pauli‐product measurements; teleportation primitives comprise Bell‐pair preparation, joint Bell‐basis measurement, and feed‐forward Pauli corrections, each executed in $O(d)$ cycles; and configurable routing regions, either statically provisioned for predictable bandwidth or dynamically reconfigurable to adapt ancilla pools to runtime congestion, ensure balanced connectivity across compute, factory, and memory segments.

  \item \textbf{Classical Compilation, Mapping \& Scheduling Layer: }
    A compiler front end transpiles high‐level Clifford+$T$ circuits into sequences of native Pauli‐based operations and lattice‐surgery primitives; mapping algorithms embed the logical‐qubit interaction graph onto the 2D patch fabric, minimizing SWAP or braiding overhead under connectivity constraints; and a scheduler allocates time slots to each operation, respecting magic‐state supply, ancilla bandwidth, and asynchronous logical‐clock domains—and resolves resource contention via critical‐path or list‐scheduling heuristics.

  \item \textbf{Qubit Memory Layout: }
    This is a layer specific to qLDPC codes introduced in \cite{hetec} \cite{Bravyi_2024} \cite{he2025extractorsqldpcarchitecturesefficient}. A high‐density logical‐qubit storage layer employs qLDPC codes (e.g.\ an \([[n,k,d]]\) gross code) to encode $k$ qubits with low overhead, supporting only Clifford‐type operations; a fault‐tolerant teleportation interface between memory and compute regions is realized by preparing and measuring Bell pairs across the surface‐code and qLDPC patches in $O(d)$ cycles; and a trade‐off analysis between data‐movement cost and qubit overhead places cold data in the qLDPC tier to minimize total patch count while keeping hot data in the surface‐code fabric for rapid gate access.  

\end{itemize}
We discuss two of the most recent heterogeneous and homogeneous architectures proposed for near-term FTQC in the following case studies.  

\begin{svgraybox}

\textbf{Case Study 4.1: HetEC \cite{hetec}}

\textbf{Logical-Qubit Fabric: }
At the heart of HetEC lies a checkerboard of planar surface-code patches, each of dimension \(d\times d\), encoding one logical qubit of distance \(d\). Nearest-neighbor stabilizer measurements enforce error correction with a pseudothreshold around 0.7\%. Clifford gates are realized via lattice surgery—merging and splitting patches—while non-Clifford \(T\) gates are executed by injecting distilled magic states into these patches. Logical error rates under surface-code operations scale as \(\sim0.03\,(p/0.01)^{(d+1)/2}\), and each Pauli-rotation requires between \(d\) and \(3d\) measurement rounds depending on its weight.

\textbf{Magic-State Factory: }
A dedicated magic-state factory sits adjacent to the compute fabric, continuously distilling \(\lvert T\rangle\) ancillas. By decoupling magic-state distillation from the main surface-code region, HetEC maintains a steady supply of high-fidelity non-Clifford resources without bloating the compute area. Distilled \(\lvert T\rangle\) states are teleported into compute patches via lattice-surgery protocols, preserving universality while containing overhead within the factory block.

\textbf{Ancilla Bus for Inter-Code Data Movement: }
Compute and memory regions communicate through a 103-qubit ancilla bus, which interfaces with a single “port” qubit on each qLDPC memory block. Within the gross code, logical automorphism gates permute any target qubit to this port in just two syndrome cycles. Thereafter, a joint \(XX\) (or \(ZZ\)) measurement over the bus teleports the qubit state fault-tolerantly between memory and compute. This bi-directional teleportation mechanism underpins all inter-code data transfers.

\textbf{HetEC Transpiler \& Scheduler: }
The HetEC transpiler accepts high-level Clifford+\(T\) circuits and performs:
\begin{enumerate}
  \item Decomposition into weighted Pauli-product rotations.
  \item Constraint-aware pruning and commuting of Clifford rotations to fit surface-code capacity.
  \item Scheduling of logical operations, data-movement calls, and in-memory Clifford executions to minimize teleportation steps and respect asynchronous syndrome clocks.
\end{enumerate}
This compilation flow balances compute-region limits against memory-region efficiency, reducing costly inter-code transfers.

\textbf{Memory layout: }
Quantum memory is provided by a \([[144,12,12]]\) qLDPC CSS gross code, encoding 12 logical qubits with a 1/24 overhead and a \(\sim0.65\%\) physical threshold. One logical qubit serves as the bus port; the remaining 11 support native Clifford gates via in-block automorphism circuits (parallel CNOT layers), each completing in \(\sim14\) syndrome cycles. By executing most Clifford operations in memory, HetEC offloads the compute fabric and reduces overall qubit pressure.

\textbf{Integration: }
By co-designing heterogeneous codes, an ancilla bus, and a specialized compiler, HetEC achieves up to a 6.42× reduction in physical qubits at the cost of a 3.43× slower logical-clock depth for benchmark circuits. 
\end{svgraybox}

\begin{svgraybox}
    \textbf{Case Study 4.2: SPARO \cite{sparo}}
    
\textbf{Logical-Qubit Fabric: }
SPARO assumes a uniform array of rotated surface-code patches, each of code distance \(d\), supporting nearest-neighbor stabilizer checks and lattice-surgery operations for both Clifford and Pauli-product measurements (PPMs).  Multi-qubit PPM errors are modeled as
\[
P_{\rm op}^{(t)} = f\bigl(n_q^{(t)},\,\ell_{\rm anc}^{(t)},\,\mathrm{op\_type}\bigr),
\]
where \(n_q^{(t)}\) is the number of qubits involved in layer \(t\), \(\ell_{\rm anc}^{(t)}\) is the ancilla-path length, and \(\mathrm{op\_type}\) distinguishes \(X\!/Z\) versus \(Y\) measurements.  Single- and two-qubit depolarizing channels are incorporated (with two-qubit error probability \(p_{\rm dep2}/16\)), and patch rotations to realign boundaries incur additional error and latency.

\textbf{Magic-State Factory: }
The baseline SPARO layout includes a single magic-state factory consuming 11 tiles to run a 15-to-1 distillation protocol, plus up to 2 ancillary tiles for \(Y\)-measurement decomposition if twist defects are unavailable.  Magic states are produced at discrete timesteps (every 11 syndrome cycles) and qubits requiring a \(T\)-gate may idle while waiting, contributing a layer-wise idle error
\[
P_{\rm idle}^{(t)} = 1 - \prod_{i=1}^{n_{\rm idle}^{(t)}} \bigl(1 - p_{i,t}\bigr),
\]
with \(p_{i,t}\) the physical idling error rate for qubit \(i\) in layer \(t\).

\textbf{Routing Region \& Ancilla Pathfinding: }
Rather than a fixed bus, SPARO embeds a routing subregion next to compute patches. For each PPM, ancilla qubits traverse this region; SPARO formulates ancilla-path minimization as a Steiner-tree problem on the routing graph, pruning candidate Steiner vertices by Manhattan distance before invoking a classical solver to yield the shortest path length \(\ell_{\rm anc}\).  Shorter \(\ell_{\rm anc}\) directly reduces both the operation error and latency.

\textbf{End-to-End PBC Pipeline: }
SPARO’s compiler proceeds in three key phases:
\begin{enumerate}
  \item \textbf{PPM Decomposition \& Pathfinding}: Translates Clifford+\(T\) circuits into weighted Pauli rotations and computes minimal ancilla paths via Steiner trees.
  \item \textbf{Operation Scheduling}: 
    Rotation scheduling minimizes basis-shift overhead by reordering \(Y\)-measurements within each layer.
    Measurement parallelization builds an interaction graph of PPMs and applies graph‑coloring to maximize concurrent measurements.

  \item \textbf{Qubit Mapping}: Performs a greedy placement of high-use qubits near factories, then refines via simulated annealing to minimize a cost function based on total Manhattan distance between interacting qubits.
\end{enumerate}

\textbf{Memory Layout: }
In SPARO, “memory” is implicit: qubits not participating in a given layer simply idle on the surface code.  The comprehensive logical-error model then computes for each layer \(t\):
\[
p_L^{(t)} = 1 - \bigl(1 - P_{\rm op}^{(t)}\bigr)\,\bigl(1 - P_{\rm idle}^{(t)}\bigr),
\quad
p_{\rm total}^L = 1 - \prod_{t=1}^{T}\bigl(1 - p_L^{(t)}\bigr),
\]
capturing failures from active PPMs, rotations, and idling.

\textbf{Integration: }
SPARO first quantifies total time and error contributions per category over all layers—\((T_{\rm total}^{\rm WaitT},E_{\rm total}^{\rm WaitT})\), \((T_{\rm total}^{\rm Rotation},E_{\rm total}^{\rm Rotation})\), and \((T_{\rm total}^{\rm Op},E_{\rm total}^{\rm Op})\).  It then estimates the marginal error-reduction yield of adding one unit of resource area to either factories or routing:
\[
\frac{\lvert\Delta E_{\rm total}^{\rm WaitT}\rvert}{A_{\rm factory}}
\quad\text{vs.}\quad
\frac{\lvert\Delta E_{\rm total}^{\rm Op} + \Delta E_{\rm total}^{\rm Rotation}\rvert}{A_{\rm routing}}
\]
and greedily allocates resources to whichever option maximizes \(\Delta p_{\rm total}^L / \Delta A\).  After each allocation step, SPARO re-runs affected compilation phases to update bottleneck statistics. 
Applying this dynamic approach under a fixed total-tile budget, SPARO achieves up to 51.11\% reduction in logical-error rate for a 433-qubit adder—outperforming both the Minimal baseline and an Intermediate static layout ($\sim30\%$ extra tiles) across multiple benchmarks.  

\end{svgraybox}

%% file: author/section5.tex
\section{Future of FTQC research}
\label{sec5}

Future work should focus on bringing improvement to the existing techniques of design automation, as well as exploring other angles on reducing the runtime for current optimization ideas in QEC implementation, and modifying the near-term FTQC architecture to accommodate multiple qubit technologies to improve the flow of design. Additionally, incorporating modifications in the modern QEC codes for accounting for defects introduced due to radiation and calibration errors in quantum computers and accelerating the decoding algorithms using Application-Specific Integrated Circuits (ASICs) and Field-Programmable Gate Arrays (FPGAs) are plausible directions for progressing towards an FTQC future.